\documentclass[12pt,preprint]{aastex}

\def\keV{{\rm\,keV}}

\def\msun{{\rm\,M_\odot}}

\def\vol#1  {{{#1}{\rm,}\ }}

\def\etal{et al.\ }

\newcount\refno
\newcount\rfno
\def\eq{$^{\the\refno\ }$\advance\refno by 1}
\def\ad{\advance\rfno by 1}

\def\clock{\count0=\time \divide\count0 by 60
     \count1=\count0 \multiply\count1 by -60 \advance\count1 by \time
     \number\count0:\ifnum\count1<10{0\number\count1}\else\number\count1\fi}

\def\myputfigure#1#2#3#4#5%
{\hskip0.03\textwidth\vskip#5pt
\makebox[0pt]{\hskip#2in
\includegraphics[width=#3\textwidth]{#1}}\vskip#4pt\hfill}

\def\Gcm2{\rm G~cm^2}

\def\beq{\begin{equation}}
\def\eeq{\end{equation}}

\def \date         {\ifcase\month \message{zero} \or
                    January \or February \or March \or April \or May \or June
                    \or July \or
                    August \or September \or October \or November \or
                    December \fi
                    \space\number\day, \number\year}

\begin{document}
\title{Mass-Temperature Relation of
Galaxy Clusters: A Theoretical Study}
\author{Niayesh Afshordi\altaffilmark{1} and Renyue Cen\altaffilmark{2}}
\altaffiltext{1} {Princeton University Observatory, Princeton
University, Princeton, NJ 08544; afshordi@astro.princeton.edu}
\accepted{ } \altaffiltext{2} {Princeton University Observatory,
Princeton University, Princeton, NJ 08544;
cen@astro.princeton.edu} 

\begin{abstract}

Combining conservation of energy throughout nearly-spherical
collapse of galaxy clusters with
the virial theorem, we derive the mass-temperature relation
for X-ray clusters of galaxies $T=CM^{2/3}$.
The normalization factor $C$
and the scatter of the relation are determined from first principles
with the additional assumption of initial Gaussian random field.
We are also able to reproduce the recently observed break
in the M-T relation at $T \sim 3 \keV$,
based on the scatter in the underlying density field for
a low density $\Lambda$CDM cosmology.
Finally, by combining observational data of
high redshift clusters with our theoretical formalism, we find a
semi-empirical
temperature-mass relation which is expected to hold
at redshifts up to unity
with less than $20\%$ error.

\end{abstract}


\section{Introduction}

The abundance of clusters of galaxies provides one of the
strongest constraints on cosmological models (Peebles, Daly, \&
Juszkiewicz 1989; Bahcall \& Cen 1992;
White, Efstathiou, \& Frenk 1993; Viana \& Liddle 1996; Eke, Cole,
\& Frenk 1996; Oukbir, Bartlett, \& Blanchard 1997; Bahcall, Fan,
\& Cen 1997; Pen 1998; Cen 1998; Henry 2000; Wu 2000) with an uncertainty
on the amplitude of density fluctuations of about 10\%
on $\sim 10h^{-1}$Mpc scale.
Theoretically it is often desirable to
translate the mass of a cluster, that is predicted by either
analytic theories such as Press-Schechter (1974) theory or
N-body simulations, to the temperature of the cluster,
which is directly observed.
Simple arguments based on virialization
density suggest that $T\propto
M^{2/3}$, where $T$ is the temperature of a cluster within a
certain radius (e.g., the virial radius)
and $M$ is the mass within the same radius.
However, the proportionality coefficient has not
been self-consistently determined from first principles, although
numerical simulations have frequently been used to calibrate the
relation (e.g., Evrard, Metzler, \& Navarro 1996, hereafter EMN;
Bryan \& Norman 1998; Thomas et al. 2001).

It is noted that the results
from different observational methods of mass measurements are not
consistent with one another
and with the simulation results (e.g.,
Horner, Mushotzky, \& Scharf 1999, hereafter HMS; Neumann, \&
Arnaud 1999; Nevalainen, Markevitch, \& Forman 2000, Finoguenov,
Reiprich, \& Bohringer 2001, hereafter FRB ).
   In general, X-ray mass estimates are about $80\%$ lower than
   the predictions of hydro-simulations.
Fig 1 compares X-ray cluster observational data with best fit line to
EMN simulation results (FRB).
   On the other hand, mass
   estimates from galaxy velocity dispersion
  seem to be consistent with simulation results (HMS).
The error in the gravitational
   lensing mass measurements is still too big to distinguish between these
   two (Hjorth, Oukbir \& van Kampen 1998).

\clearpage
\myputfigure{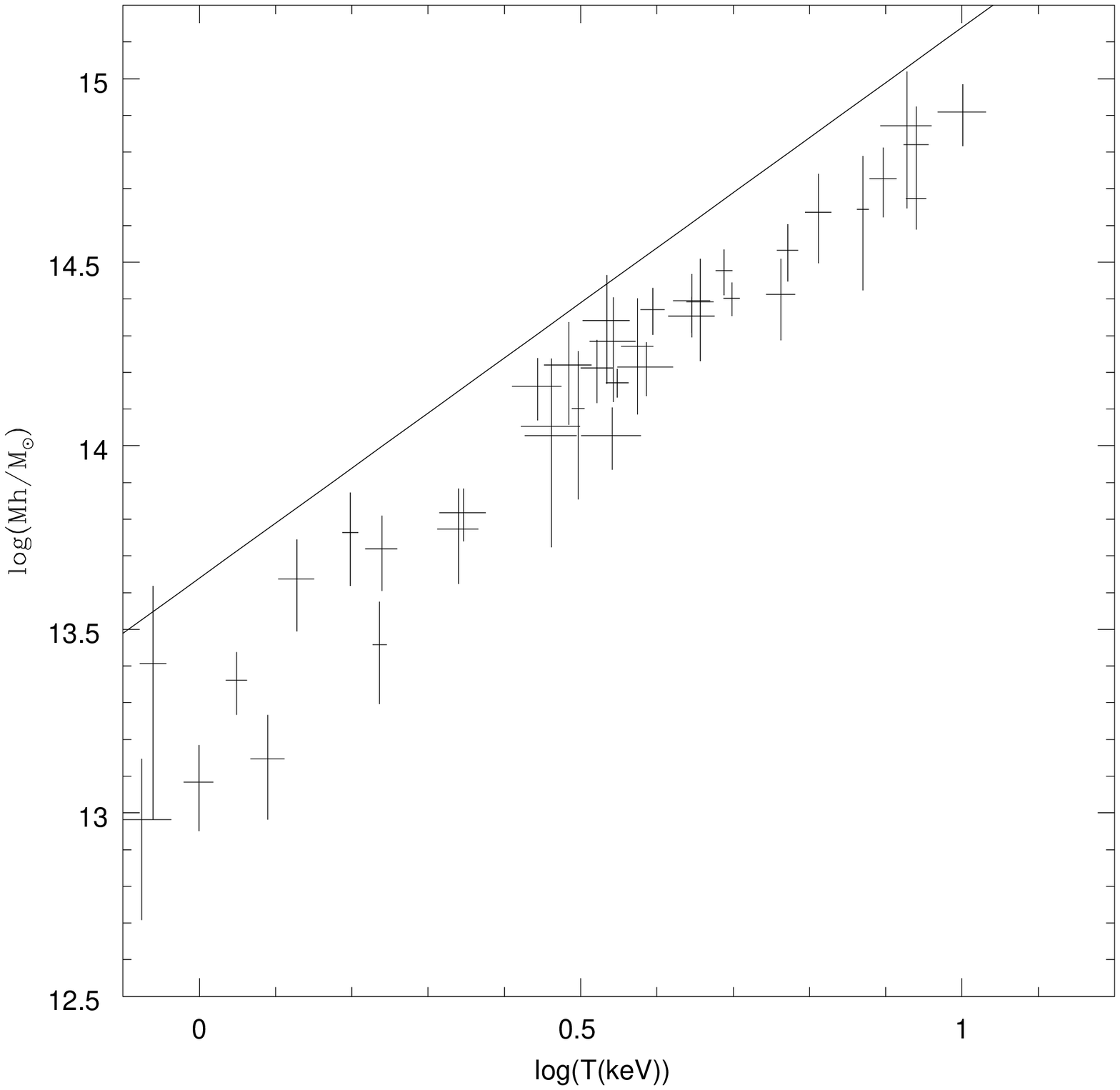}{6.0}{0.80}{-10}{-10}
\vspace{-1.0cm}
\figcaption{Observational
mass-temperature data based on temperatures of X-ray clusters
(crosses with $1\sigma$ error bars) vs. the best fit line
to EMN simulation results (FRB).
\label{fig1}}
\vspace{\baselineskip}
\clearpage

   Another recent observational
 finding is the possible existence of a break in the $T-M$
   relation.
   By use of resolved temperature profile of X-ray clusters
   observed by ASCA,
   FRB have investigated $T-M$ relation in the low-mass end and
   find that $M\propto T^{\sim 2}$,
   compared to
   $M\propto T^{\sim 3/2}$ at the high mass end.
   Suggestions
   have been made to explain this behavior by attributing it the effect of
   formation redshift (FRB), cooling (Muanwong \etal 2001)
   and heating (Bialek, Evrard \& Mohr 2000) processes.

    In this paper we use conservation of energy for an almost spherically
    collapsing region to derive the M-T relation. In \S 2
    we find the initial and final energy of the cluster.
    \S 3 constrains various factors which enter the
    normalization of M-T relation, via statistical methods,
    simulation and observational input. \S 4 considers
    predictions of our model and comparison with observational
    and simulation results. In \S 5, we discuss the
    limitations of our approach and justify some of the
    approximations. \S 6 concludes the paper.

\section{Conservation of Energy}
\subsection{Initial Kinetic and Potential Energy of a Proto-Cluster}

  We begin by deriving the kinetic energy of the proto-cluster.
 We can write velocity as a
 function of gravitational potential $\phi_i$ (e.g. Padmanabhan 1993):
 \begin{equation}
    {\mathbf v} = H_i {\mathbf x}-\frac{2}{3H_i}{\mathbf \nabla}\phi_i,
 \end{equation}
 at the initial time
 in the linear regime,
 where $H_i$ is the Hubble constant at the
 initial time. There is a small dependence on
 the initial density parameter $\Omega_i$ in equation (1)
 which we ignore since initially it is very close to unity and
 the difference will be in second order terms that we ignore for
 the proto-cluster.
 Then the kinetic energy is given by:
 \begin{equation}
 K_i = \frac{1}{2}\int\rho v^2 d^3x =
\frac{1}{2}\rho_{i}\int(1+\delta_i)|H_i{\mathbf x}-\frac{2}{3H_i}{\mathbf \nabla}\phi_i|^2
d^3x.
 \end{equation}
 Keeping the terms up to the linear order we obtain
\begin{equation}
 K_i = \frac{1}{2}\rho_i \int(H_i^2 x^2 +\frac{2}{3}\nabla^2\phi_i
 x^2 - \frac{4}{3} {\mathbf x}.{\mathbf \nabla}\phi_i) d^3x.
 \end{equation}
 In deriving equation (3), we have used the Poisson equation:
 \begin{equation}
 \nabla^2 \phi_i = 4\pi G \rho_i \delta_i,
 \end{equation}
 to substitute for $\delta_i$,
 where $\rho_i$ is the initial mean density of the universe.
 For equation (3) we then use Gauss theorem to make
 the third term similar to the second, at the expense of a surface
 term:
 \begin{equation}
 K_i = \frac{1}{2}\rho_i\int(H_i^2
 x^2+\frac{4}{3}x^2\nabla^2\phi_i)d^3x -\frac{1}{3}\rho_i\oint x^2
 {\mathbf \nabla}\phi_i.d{\mathbf a}.
 \end{equation}
 Assuming that the deviations from spherical symmetry is not
 important at the boundary of the proto-cluster,
 we find $\nabla\phi_i$ in
 the second term in equation (5)
 as a function of $\delta_i$:
 \begin{equation}
 {\mathbf \nabla}\phi_i= \hat{{\mathbf r}}\frac{G \delta M}{R_i^2} = \hat{{\mathbf r}}\frac{G
 \rho_i}{R_i^2}\int \delta_i d^3x,
 \end{equation}
 where $G$ is the gravitational constant and
 $R_i$ is the boundary radius of the initial proto-cluster,
which leads to
 \begin{equation}
  K_i = \frac{4\pi G \rho_i^2}{3\Omega_i}\int [
    (1+2\delta_i)x^2-R_i^2 \delta_i] d^3x,
 \end{equation}
 where we have used equation (4) to substitute back for $\phi_i$ and also the
 definition of $\Omega_i \equiv 8\pi G \rho_i/(3 H_i^2)$.

 Let us now find an expression for the gravitational potential energy
 of the proto-cluster. Using its definition we have
\begin{equation}
 U_i =
 -\frac{G\rho_i^2}{2}\int\int[\frac{(1+\delta_i(x_1))(1+\delta_i(x_2))}{|x_1-x_2|}]d^3x_1
 d^3x_2.
 \end{equation}
Keeping the terms to the first order and using the symmetry under
interchange of $x_1$ and $x_2$, we arrive at
\begin{equation}
U_i =
-\frac{G\rho_i^2}{2}\int(1+2\delta_i(x_1))d^3x_1\int\frac{d^3x_2}{|x_1-x_2|},
\end{equation}
which, taking the second integral in a spherical volume, gives
\begin{equation}
U_i = -\frac{4\pi G}{3}\rho_i^2\int
(1+2\delta_i)(\frac{3R_i^2-x^2}{4})d^3x.
\end{equation}

\noindent Adding equations 7 and 10 gives the total initial
energy\footnote{It is easy to see that $1-\Omega_i = O(\delta_i)$ and so 
we have neglected it in the first order terms.}
\begin{equation}
E_i = \frac{4\pi
G}{3}\rho_i^2[\frac{4\pi}{5}(1-\Omega_i)R_i^5-\frac{5}{2}\int\delta_i(x)(R_i^2-x^2)d^3x],
\end{equation}
to the first order. Defining $\tilde{x}$, $\tilde{\delta}_i$ and
$B$ as:
\begin{equation}
\tilde{x}\equiv\frac{x}{R_i},\, \tilde{\delta}_i \equiv \delta_i +
\frac{3}{5}(\Omega_i-1), B \equiv \int_0^1\tilde{\delta}_i(\tilde{x})
 (1-\tilde{x}^2)d^3\tilde{x},
\end{equation}
 equation (11) is simplified:
 \begin{equation}
 E_i = -\frac{10\pi G}{3} \rho_i^2 R_i^5 B.
 \end{equation}
 The integral in the definition of $B$ is in fact a three dimensional integral and
 the limits denote that the integration domain is the unit sphere.
 Note that $\tilde{\delta}_i$  can be considered as the density
perturbation
 to a flat (i.e. $\Omega_{tot} = 1$) universe for which energy vanishes.
 Another aspect of this statement is that both terms in the
 definition of $\tilde{\delta}_i$ scale as $a$ in the early (matter
 dominated) universe and so does $\tilde{\delta}_i$ itself.
 For a flat universe the first term
 dominates at high redshift since the second term
 scales as $a^3$.

\subsection{Energy of a Virialized Cluster}

According to the virial theorem, the sum of the total energy $E_f$
of a
virialized cluster and its kinetic energy $K_f$
vanishes. However,
non-vanishing pressure at the boundary of
the cluster can significantly modify the virial relation.
Integrating the equation of hydrostatic equilibrium, we have
\begin{equation}
 K_f+E_f = 3P_{ext}V,
\end{equation}
 where $P_{ext}$ is the pressure on the outer boundary of
 virialized region (i.e. virial radius) and V is the volume.
 For now we assume that the surface term is related to
 the final potential energy $U_f$ by
\begin{equation}
3P_{ext}V = -\nu U_f.
\end{equation}
 We will consider the coefficient $\nu$ and its possible mass dependence
 in \S 3.4.
For a system of fully ionized gas plus dark matter, the virial
relation (14) with equation (15) leads to
\begin{equation}
-(\frac{1+\nu}{1-\nu})E_f = K_f = \frac{3}{2} M_{DM} \sigma_v^2+
\frac{3M_{gas} k T}{2\mu m_p},
\end{equation}
 where $\sigma_v$ is the
 mass-weighted mean one-dimensional velocity dispersion of dark
 matter particles,
 $M_{DM}$ is the total dark matter mass,
 $k$ is the Boltzmann constant,
 $\mu = 0.59$ is the mean molecular weight and
 $m_p$ is the proton mass.
 Assuming that the ratio of gas to dark matter mass
 in the cluster is the same as that of the
 universe as a whole
 and $f$ is the fraction of the baryonic matter in the hot gas,
 we get
 \begin{equation}
 K_f = \frac{3\beta_{spec} M k T}{2\mu m_p}[1 +
 (f \beta_{spec}^{-1}-1)\frac{\Omega_b}{\Omega_m}].
\end{equation}
 with $ \beta_{spec} \equiv \sigma_v^2/(kT/\mu m_p)$.
 Hydrodynamic simulations indicate that $\beta_{spec} \simeq 1$.
 For simplicity we define $\tilde{\beta}_{spec}$ as
 \begin{equation}
\tilde{\beta}_{spec} = \beta_{spec} [1 +
 (f \beta_{spec}^{-1}-1)\frac{\Omega_b}{\Omega_m}].
 \end{equation}
So equation (16) reduces to:
\begin{equation}
K_f = \frac{3\tilde{\beta}_{spec} M k T}{2\mu m_p}
\end{equation}

\noindent
Assuming energy conservation (i.e., $E_i=E_f$)
and combining this result with equations
(13, 16) lead to the temperature as a function
 of initial density distribution:
 \begin{equation}
  k T = \frac{5 \mu m_p}{8 \pi \tilde{\beta}_{spec}} (\frac{1+\nu}{1-\nu}) \, H_i^2 R_i^2
  B.
  \end{equation}
\noindent
In the next subsection \S 2.3
we will find an expression for
$H_i^2 R_i^2$ in terms of
cluster mass $M$
and the initial density fluctuation spectrum.

\subsection{Virialization Time}
 Defining $e$ as the energy of a test particle with unit mass,
 which is at the boundary of the cluster $R_i$ of mass $M$ initially:
 \begin{equation}
 e = \frac{\mathbf{v}^2_i}{2}-\frac{GM}{R_i}.
 \end{equation}
 The collapse time $t$ can be written as
 \begin{equation}
 t = \frac{2\pi G M}{(-2e)^{\frac{3}{2}}}.
 \end{equation}
Following the top-hat model, we assume that the collapse time
of the particle is
approximately the same as the time necessary for the particle to
be virialized.
Choosing $t$ to be
the time of observation and assuming the mass $M$, interior to
the test particle is virialized at $t$,
and by combing equations (1, 6, 22),
we find $e$ as function of initial density distribution
and relate it to the collapse time:
 \begin{equation}
 -2e = \frac{5}{4 \pi}H_i^2 R_i^2\int_0^1 \tilde{\delta}_i(\tilde{x})
 d^3\tilde{x} = (\frac{2\pi G M}{t})^{\frac{2}{3}}.
 \end{equation}
Using $M = \frac{4}{3} \pi R_i^3 \rho_i$ and the
Friedmann equations we obtain
 \begin{equation}
A \equiv \int_0^1 \tilde{\delta}_i(\tilde{x})
 d^3\tilde{x} = \frac{2}{5}(\frac{3\pi^4}{t^2
 G \rho_i})^{\frac{1}{3}}.
 \end{equation}

\subsection{M-T Relation}

   Combining equations (20) and (13,24), we arrive at the cluster
   temperature-mass
   relation:
\begin{equation}
  kT = (\frac{\mu m_p}{2\tilde{\beta}_{spec}})(\frac{1+\nu}{1-\nu})(\frac{2\pi G
  M}{t})^{\frac{2}{3}}(\frac{B}{A}).
\end{equation}
  Notice that although $B$ and $A$ are both functions of the
  initial moment $t_i$, since both are proportional to the scale
  factor $a$, the ratio is a constant;
  the derived T-M relation (equation 25)
  does not depend on the adopted initial time,
  as expected.
  As a specific example, in the spherical top-hat
  model in which the density contrast is assumed to be constant,
  this ratio $B/A$ is $\frac{2}{5}$.
  Let us gather all the unknown dimensionless factors in $\tilde{Q}$:
  \begin{equation}
  \tilde{Q} \equiv
(\frac{\tilde{\beta}_{spec}}{0.9})^{-1}(\frac{1+\nu}{1-\nu})(\frac{B}{A})
  (Ht)^{-2/3}.
  \end{equation}
   Then, inserting in the numerical values, equation (25) reduces to:
   \begin{equation}
   kT = (6.62 \, \keV)\tilde{Q}(\frac{M}{10^{15} h^{-1}
   M_{\odot}})^{2/3},
   \end{equation}
   or equivalently:
   \begin{equation}
   M = 5.88\times 10^{13} \tilde{Q}^{-1.5}(\frac{kT}{1 \,\, \keV})^{1.5}
h^{-1} M_{\odot}
   \end{equation}
   where $H = 100 h~$km/s/Mpc is the Hubble constant.
   This result can be compared with the EMN simulation results:
   \begin{equation}
   M_{200} = (4.42 \pm 0.56)\times 10^{13} (\frac{kT}{1\,\, \keV})^{1.5}
   h^{-1} M_{\odot}.
   \end{equation}
To convert the $M_{500}$ masses of EMN to $M_{200}$
    we have used the observed scaling
   of the mass with density contrast $M_{\delta} \propto \delta ^{-0.266}$
   (HMS), which is consistent with the NFW profile
   (Navarro, Frenk, \& White 1997)
   for simulated dark matter halos as well as observations
   (e.g., Tyson, Kochanski, \& dell'Antonio 1998),
   in the relevant range of radius.

\section{Numerical Factors, Scatter and Uncertainties}

In this section we try to use different analytical methods
as well as the results of dark matter simulations and observed gas properties, available in the
 literature, to constrain the numerical factors which appear in
 $\tilde{Q}$ (equation 26).
\subsection{$\beta_{spec}$ and the gas fraction}

 So far
 we have made no particular assumption about the gas dynamics
 or its history,
 \footnote{With the exception of any
 heating/cooling of the gas being negligible with respect to the gravitational
 energy of the cluster.} and so we are going to rely on the available
observational results to
 constrain gas properties.
   $\beta_{spec}$ is defined as the ratio
   of kinetic energy per unit mass of dark matter
 to the thermal energy of
 gas particles. This ratio is typically of the order of unity,
 though different observational and theoretical methods lead
 to different values.

 The hydrodynamic simulation results usually point
   to a larger value of $\beta_{spec}$.
   For example, Thomas \etal (2001) find $\beta_{spec} = 0.94 \pm 0.03$.
   On the other hand,
   observational data point to a slightly lower value of $\beta_{spec}$.
   Observationally
   there is yet no direct way of accurately measuring the
   velocity dispersion of dark matter particles
   in the cluster and one is required
   to assume that the velocity
   distribution of galaxies follows that of dark matter
   or adopt a velocity bias.
   Under the assumption of no velocity bias
   Girardi \etal 1998 find it to be $0.88 \pm 0.04$.
   Girardi \etal 2000
   study $\beta_{spec}$ for a sample of high redshift
   clusters and do not find any evidence for redshift dependence.
   From the theoretical point of view, the actual value of $\beta_{spec}$
   might be substantially different
   from the observed number,
   because both the velocity and
   density of galaxies do not necessarily follow
   those of the dark matter,
   which could have resulted in some non-negligible
   selection effects.
 Unknown sources of heating such as due to
    gravitational energy on small scales
    which is often substantially underestimated in simulations
    due to limited resolutions
   or baryonic processes like supernova feedback
   may affect the value of $\beta_{spec}$ as well.

  Hydrodynamic simulations show that only a
 small fraction of the baryons contribute to galaxy formation in large
 clusters (e.g., Blanton \etal 2000)
 and so $f$ is close to one. Observationally
 we quote Bryan (2000) who compiled different observations for the
 cluster mass fraction in gas and galactic component with
 \begin{equation}
   f = 1- 0.26(T/10\keV)^{-0.35},
 \end{equation}
  albeit with a large scatter in the relation.
  Inserting equation (30) into
  equation (18), we see that the correction to $\beta_{spec}\sim 0.9$
  is less than $5\%$
  for all of feasible cosmological models which are dominated by
  non-baryonic dark matter.
  In what follows, unless mentioned otherwise, we adopt the value
  $\tilde{\beta}_{spec} = 0.9$ and absorb any correction into the overall
  normalization of the T-M relation.

\subsection{$B/A$: Single Central Peak Approximation and Freeze-Out time}
  In the original top-hat approximation (Gunn \& Gott 1972) which
  has been extensively used in the literature, the initial density
  distribution is assumed to be constant inside $R_i$,
  which leads
  to $B/A=2/5$.
  Shapiro, Iliev, \& Raga (1999) and Iliev \& Shapiro (2001)
  have extended the original treatment of top-hat spherical
  perturbation to a more self-consistent case with a 
  trancated isothermal sphere final density distribution
  including the surface pressure term (see \S 3.4).
  Here we consider the general case with an arbitrary density profile
  of a single density peak.
  But before going further,
  let us separate out the term due to space curvature
  in the definition of $B/A$. Using the definitions of $B$ and $A$ in
  equations
  (12) and (24), and Friedmann equations to insert for $\rho_i$ in terms
  of the present day cosmological parameters, we get:
  \begin{equation}
     \frac{B}{A} =
\frac{b}{a}+(\frac{b}{a}-\frac{2}{5})(1-\Omega_m-\Omega_{\Lambda})(\frac{Ht}{\Omega_m
\pi})^{\frac{2}{3}},
  \end{equation}
  where
  $\Omega_m$ and $\Omega_\Lambda$ are the density parameters due to
  non relativistic matter and cosmological constant, respectively,
  at the time of observation,
  $a$ and $b$ are\footnote{$a$ should not to be confused with the
cosmological
scale
factor.} the same as $A$ and $B$ with $\tilde{\delta_i}$ replaced by $\delta_i$ in their
  definitions, equations (24) and (12).

  Assuming that the initial density profile has a single,
  spherically symmetric peak and
  assuming a power law for the initial linear correlation function at
  the cluster scale, we can replace $\delta(x)$
  by $\frac{\xi(x)}{\delta{(0)}}$,
  \begin{equation}
  \xi(x)=  (\frac{r_{0i}}{x})^{3+n},
  \end{equation}
where $n$ is the index of the density power spectrum (Peebles 1981)
and $r_{0i}$ is the correlation length.
  This gives
  \begin{equation}
\frac{B}{A} = (\frac{1}{1-\frac{n}{2}})[1+ \frac{
  (3+n)(1-\Omega_m-\Omega_{\Lambda})}{5}(\frac{Ht}{\pi
\Omega_{m}})^{\frac{2}{3}}],
  \end{equation}
   where all the quantities are evaluated at the age
   of the observed cluster.
   Note that for models of interest the physically plausible range
   for $n$ is $(-3,0)$.
    One can see that the second term in equation (33)
    is indeed proportional to
 $t^{\frac{2}{3}}$ and so for an open universe it dominates for large time
 (after curvature domination).
 Noting that in equation (25), the temperature is proportional to $t^{-\frac{2}{3}}\frac{B}{A}$, so, when the
 second term dominates, the T-M relation will no longer
 evolve with time.
 This indicates that in an open
 universe the cluster formation freezes out
 after a certain time. The presence of
 freeze-out time is independent of the central peak
 approximation since the ratio $b/a$ only depends on the statistics of the initial fluctuations
 at high redshifts where there is
 very little dependence on cosmology.
 It is interesting to note that in the case $n=-3$,
 the ratio has no dependence on cosmology and there is no
 freeze-out even in low density universes.
 This is an interesting case where linear theory does not apply,
 because all scales become nonlinear at the same time
 and {\it the universe is inhomogeneous on all scales}.

 Voit (2000) uses a different method to obtain exactly the
 same result. As we argue next, both treatments ignore cluster mergers.

 \subsection{$B/A$: Multiple Peaks and Scatter}

 The single peak approximation discussed
 in \S 3.2 ignores the presence of
 other peaks in the initial density distribution.
In hierarchical structure formation models,
the mass of a cluster grows with
 time through mergers
 as well as accretion.
 This means that multiple peaks
 may be present within $R_i$
 and suppress the effect of the central peak.

  Assuming Gaussian statistics for initial density fluctuations, we can find the
 statistics of $b/a$. Note that using equation (24), we can fix the value
 $A$ (and hence $a$) for a given mass and virialization time.
 So the problem reduces
 to finding the statistics of $b$ (or $B$) for fixed $a$.
 Under the assumption of a power law spectrum (see Appendix A) calculations give
 \begin{equation}
   \frac{<b>}{a} = \frac{4(1-n)}{(n-5)(n-2)},
 \end{equation}
 with
 \begin{equation}
 \Delta{b} = \frac{16 \pi 2^{-n/2}}{(5-n)(2-n)}[\frac{n+3}{n(7-n)(n-3)}]^{\frac{1}{2}}(\frac{r_{0i}}{R_i})^{\frac{n+3}{2}},
 \end{equation}
 which, inserting into equation (31), yields
 \begin{eqnarray}
   <\frac{B}{A}> &=&
\frac{4(1-n)}{(n-5)(n-2)}[1-\frac{n(n+3)}{10(1-n)}(1-\Omega_m-\Omega_{\Lambda})
(\frac{Ht}{\pi\Omega_m})^{\frac{2}{3}}], \\
    \frac{\Delta B}{A} &=& \tau(n)(\frac{M}{M_{0L}})^{-\frac{n+3}{6}}
D^{-1}(t) (\frac{\sqrt{\Omega_m} Ht}{\pi})^{\frac{2}{3}},
 \end{eqnarray}
 where
 $D(t)$ is the growth factor of linear perturbations, normalized to $(1+z)^{-1}$ for large redshift, and
 \begin{eqnarray}
  M_{0L} = \frac{4}{3} \pi \rho_0 r^3_{0L}, \\
  \xi_{L}(r) =  (\frac{r_{0L}}{r})^{n+3}, \nonumber
 \end{eqnarray}
  where $\xi_L(r)$ is the linearly evolved correlation function
  at the present time with $r_{0L}$ being
  the correlation length, and
 \begin{equation}
 \tau(n) \equiv \frac{20 \times 2^{-n/2}}{(5-n)(2-n)}[\frac{n+3}{n(7-n)(n-3)}]^{\frac{1}{2}}.
 \end{equation}

  Fig 2 shows the dependence of $B/A$ on $n$. The upper dotted
  curve shows the result of the single
 peak approximation (equation 33),
 while three lower curves show
 the multiple peak calculation described above (equation 36) and
 its $\pm 1\sigma$ dispersion (equation 37).
 All the curves are for an Einstein-de Sitter
 universe and the dispersion is calculated for mass $10 M_{0L}$.
 Numerically, for $r_{0L}=5h^{-1}$Mpc we have
 $M_{0L}=1.4\times 10^{14}\Omega_mh^{-1}\msun$,
 resulting in $10M_{0L}=4.3\times 10^{14}h^{-1}\msun$ for $\Omega_m=0.3$.
 We note that
 $n \sim -3$ the density distribution is dominated by the central peak
 and corresponds to the top-hat case,
 and so two methods
 give similar results.
 Interestingly,
 as $n$ approaches zero,
 small peaks dominate
 and the distribution becomes close to homogeneous (top hat) on large scales.
 This implies that clusters
 undergo a large number of mergers for large values of $n$.
 Interestingly, in this case the ratio $B/A$ again
 approaches $2/5$, the value for the top-hat case.
 We will use the multiple-peak approximation in our subsequent calculations.

    It is worth mentioning that the ratio of the cosmology dependent term in
    the average value of $<B/A>$ to the constant
    term, in the multiple-peak calculation, is small. For example, for $\Omega_m = 0.3$ and $n =
    -1.5$, this ratio is about $0.07$. This implies that the
    freeze-out time is large comparing to the current age of the universe
    for feasible open cosmological models and consequently $<B/A>$ largely
    is determined by the spectral index of the underlying linear power
    spectrum $n$.

\clearpage
\myputfigure{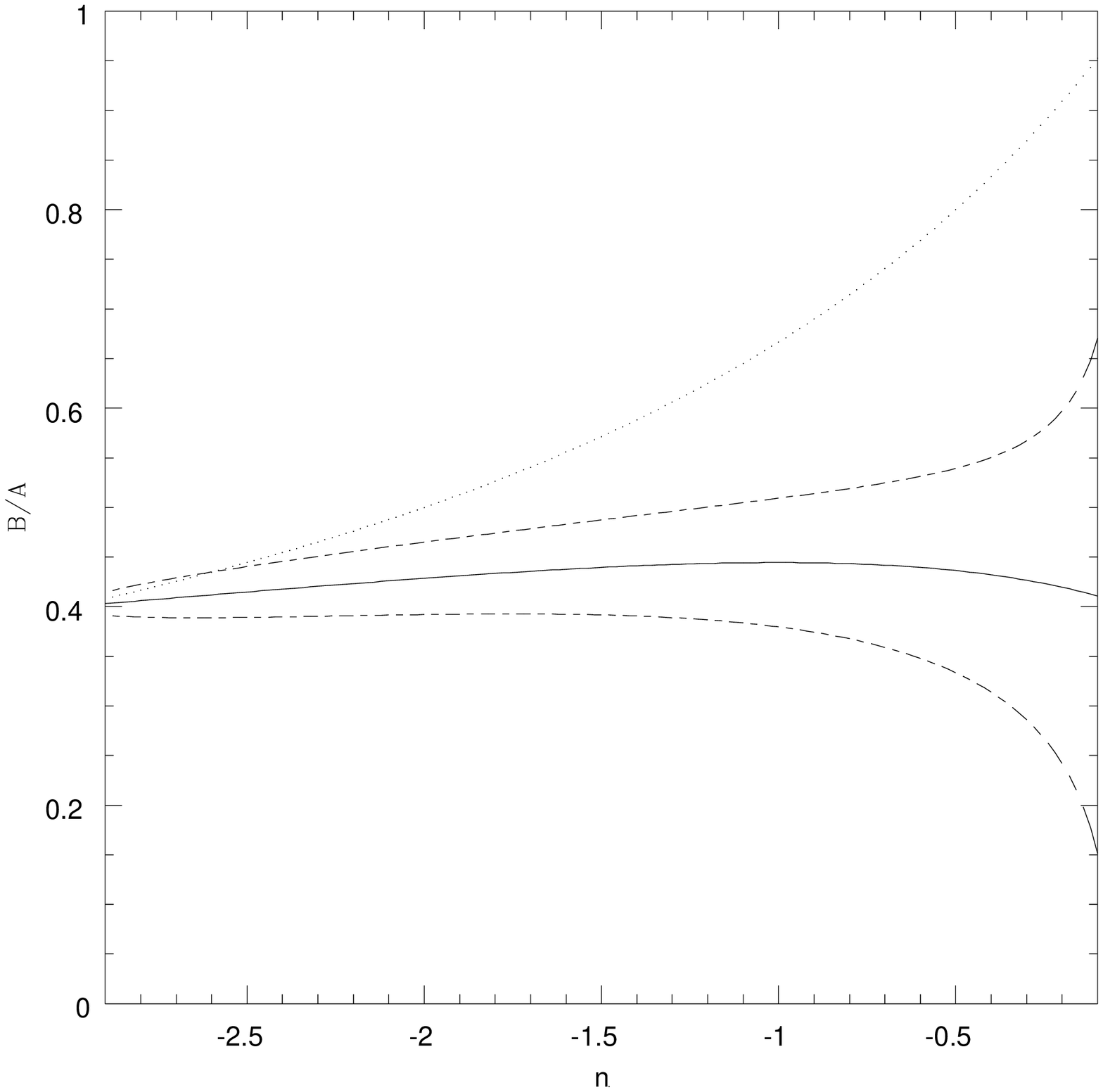}{6.0}{0.80}{-10}{-10} \vspace{-1.0cm}
\figcaption{The solid curve shows B/A vs. the power index and
dashed curves are its $\pm1\sigma$ variance for $M=10 \, M_{0L}$
(see the text) for an Einstein-de Sitter universe. The dotted
curve is the result of single peak approximation (equation 33).
\label{fig2}} \vspace{\baselineskip}
\clearpage

 \subsection{$\nu$: Surface Term and its Dependence on the final equilibrium density profile}

 As discussed in \S 2,
 corrections to the virial relation due to finite surface pressure
 changes the M-T relation (equation 25).
  Shapiro, Iliev, \& Raga (1999) and Iliev \& Shapiro (2001)
  have previously also taken into account the surface pressure term
  in their treatment of the trancated isothermal sphere equilibrium structure
  with a top-hat initial density perturbation
  and have found results in good agreement with simulations.
 Voit (2000) uses NFW profile for the final density
 distribution to constrain the extra factor
 and finds that for typical concentration parameters
 $c \equiv r_{200}/r_{s}\sim 5$, $(\nu+1)/(\nu-1)$ is $\sim 2$.
We will investigate this correction, $\nu$,
for a given concentration parameter $c$.
Let us assume that the density profile is given by:
\begin{equation}
 \rho(r) = \rho_s f(r/r_s),
\end{equation}
where $\rho_s$ is a characteristic density, $r_s$ the scale
radius, and $f$ is the density profile.
  For the NFW profile
 \begin{equation}
 f_{NFW}(x) = \frac{1}{x(1+x)^2},
 \end{equation}
Moore et al. (2000) have used simulations with higher resolutions
to show that the central density profile is steeper than
 the one already probed by low-resolution simulations such as
 those used by NFW, yielding the Moore profile
 \begin{equation}
 f_{M}(x) = \frac{1}{x^{1.5}(1+x^{1.5})}.
 \end{equation}
 For a given $f(x)$, the mass of the cluster is:
 \begin{equation}
 M = 4\pi \rho_s r^3_s g(x),\, g(x) = \int_0^c f(x) x^2 dx.
 \end{equation}
The gravitational energy of the cluster is given by:
\begin{equation}
 U = -\int \frac{G M dM}{r} = -16 \pi^2 G \rho^2_s r^5_s \int_0^c f(x) g(x) x dx.
\end{equation}
 To find the surface pressure, we integrate the equation of hydrostatic
 equilibrium\footnote{This is of course valid in the case of isotropic velocity
 dispersion profile. As an approximation, we are going to neglect any correction due to
 this possible anisotropy}
\begin{equation}
 {\mathbf \nabla}P = \rho {\mathbf g} = -\frac{GM\rho {\mathbf r}}{r^3}.
\end{equation}
 This leads to
\begin{equation}
 P_{ext} = 4 \pi G \rho^2_s r^2_s \int_c^{\infty} f(x)g(x)x^{-2}dx,
\end{equation}
which, by the definition of $\nu$ (equation 15), gives:
\begin{equation}
 \nu(c,f) \equiv -\frac{3P_{ext}V}{U}= \frac{c^3\int_c^{\infty} f(x)g(x)x^{-2}dx}{\int_0^c f(x) g(x) x dx}.
\end{equation}
Note that $\nu(c,f)$ is a function of both $c$ and density profile $f$.

If we define $Q(c,f)$ as
\begin{equation}
 Q(c,f) \equiv (\frac{1+\nu}{1-\nu}) y = (\frac{1+\nu}{1-\nu}) \frac{B}{A(Ht)^{\frac{2}{3}}},
\end{equation}

\noindent equation (25) can be written as:
\begin{equation}
 kT = (\frac{\mu m_p}{2\tilde{\beta}_{spec}})(2\pi G H M)^{\frac{2}{3}}Q(c,f).
\end{equation}
 or in numerical terms:
\begin{equation}
 kT = (6.62\,\keV) Q(c,f) (\frac{M}{10^{15} h^{-1} M_{\odot}})^{2/3}
\end{equation}
 for $\tilde{\beta}_{spec} = 0.9$. 
 Note that with this choice of $\tilde{\beta}_{spec}$, the definition
of $Q$ is equivalent to that of $\tilde{Q}$ in equation (26).

\clearpage
\myputfigure{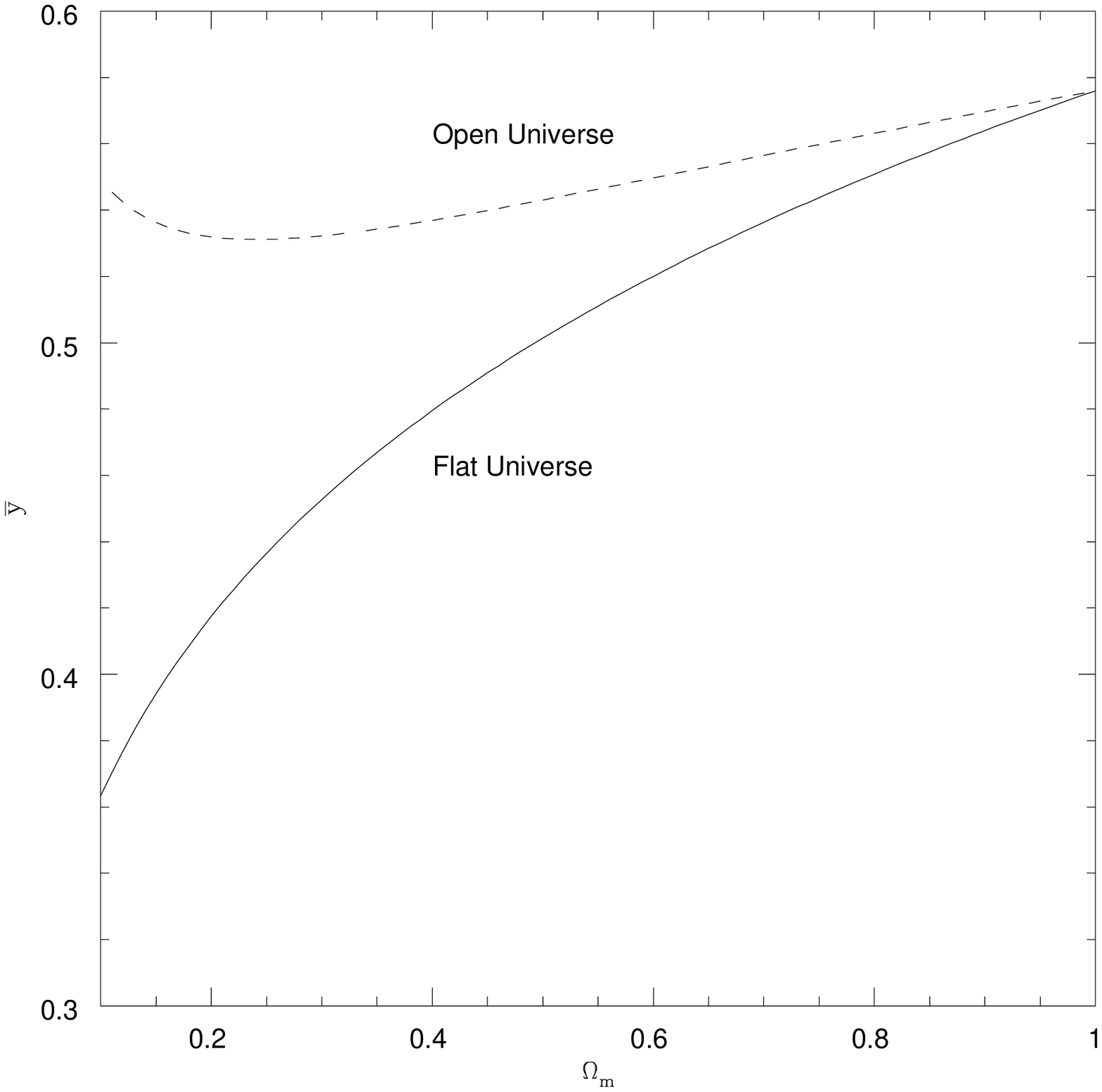}{6.0}{0.80}{-10}{-10} \vspace{-1.0cm}
\figcaption{Average value of $y$ vs. $\Omega_m$ for flat and open
cosmologies, assuming that energy is conserved. \label{fig3a}}
\vspace{\baselineskip}
\clearpage

Fig (3) shows the average value of $y$ for different cosmologies,
using equations 36 and 51, for $n =-1.5$.

\clearpage
\myputfigure{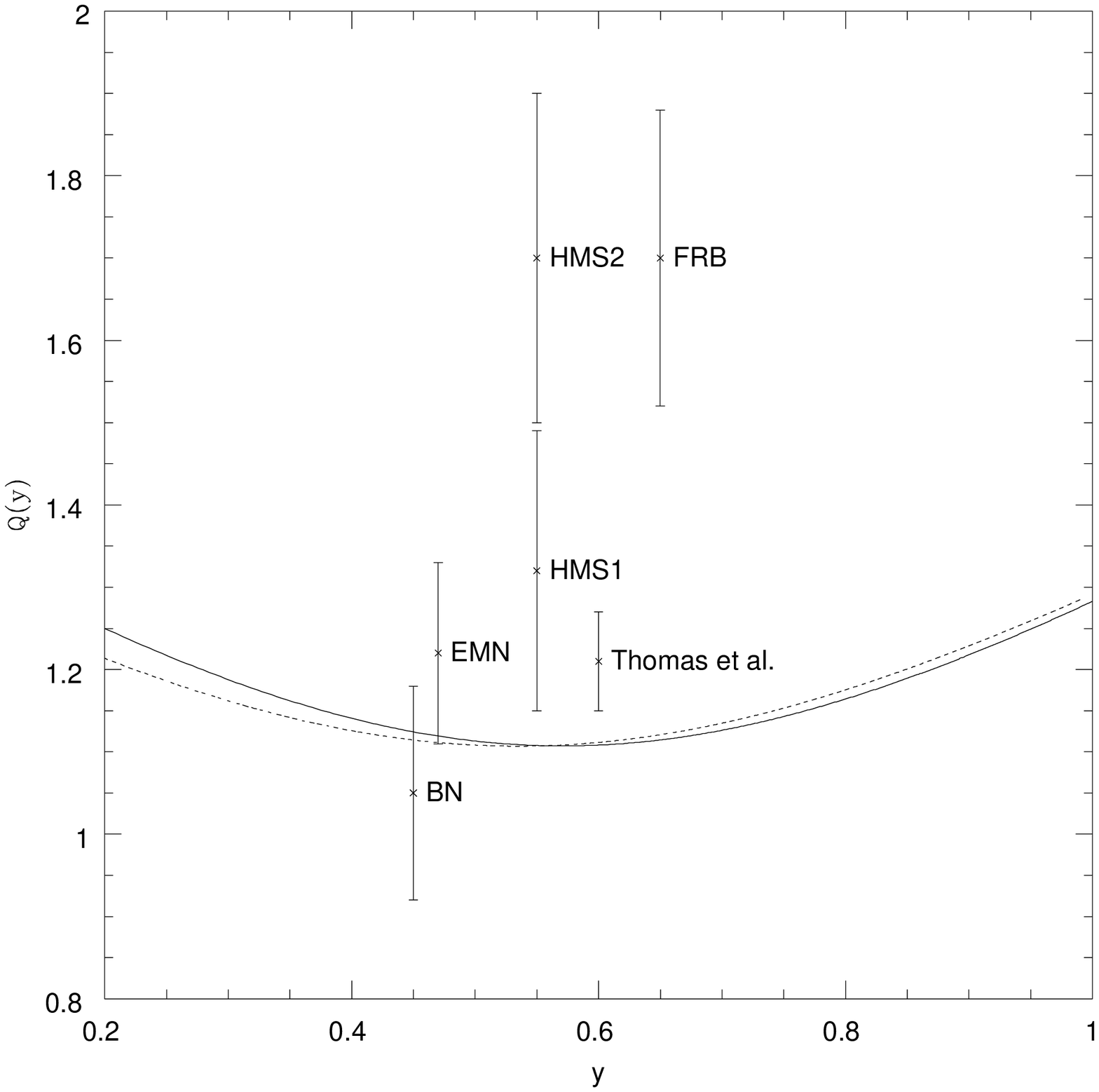}{6.0}{0.80}{-10}{-10} \vspace{-1.0cm}
\figcaption{Normalization $Q$ vs. $y$, comparison of different
results. The solid and dashed curves are for the Moore and NFW
profiles respectively. For description of different data points
see Table 1. The horizontal position of error bars is arbitrary.
\label{fig3b}} \vspace{\baselineskip}
\clearpage

 Since all three parameters,
 $\nu(c,f)$,
 $y(c,f)$ and
 $Q(c,f)$,
 are functions of both $c$ and $f$,
 one can express any one of them as a function of another, for fixed $f$.
 Fig (4) shows normalization factor $Q$ as functions of $y$. 
 The dashed curves are for the NFW profile
 and the solid curves
 for the Moore et al. (2000) profile. 
 The x's with error bars
 show various simulation and observational results (see Table 1).
 Note that even if we relax the conservation of
 energy, one can still use $Q(y)$ to find the T-M relation,
 using the value $y$ obtained from its definition, equation (49), with the
 corrected energy.

 An important feature of the behavior of $Q(y)$ is the presence of a
 minimum in or close
 to the region of physical interest. As a result, $Q(y)$ has very weak dependence on
 the history of the cluster, for example, the largest
 variation in $Q$ is about $3\% $. This is probably why simulations do not show significant
 cosmology dependence
 $\sim 5 \%$ (e.g. EMN, Mathiesen 2000).

Another way of stating this property is that the heat capacity of
the cluster is very small. It is well known that the heat capacity
of gravitationally bound systems like stars is negative. Yet we
know that if non-gravitating gas is bound by an external pressure,
its heat capacity is positive. In the case of clusters, the
interplay of external accretion pressure and gravitational binding
energy causes it to vanish. It is only after the freeze-out time
in an extreme open universe (low $\Omega_m$, no cosmological
constant) where the heat capacity becomes negative similar to an
ordinary gravitationally bound system.

\subsection{Concentration Parameter $c$}

We point out that, by using the conservation of energy, one can 
constrain the concentration parameter and subsequently the surface
correction for a given density profile. A typical density profile
is specified by two parameters: a characteristic density $\rho_s$
and scale radius $r_s$. If we know mass (e.g. $M_{200}$) and total
energy of the cluster, we can fix these two parameters. The
concentration parameter is then fixed by $\rho_s$ and the critical
density of the universe. 
To show the precedure, let us re-derive the T-M
relation for a known density profile. Combining equations (16) and
(19)
 gives:
\begin{equation}
kT = -\frac{2\mu
m_p}{3\tilde{\beta}_{spec}M}(\frac{1+\nu}{1-\nu})E_f.
\end{equation}
 Note that equation (51) only depends on the properties of the virialized
cluster and is independent of its history. Defining $y$ as
\begin{equation}
  y \equiv \frac{4E}{3M}(2\pi G M H_0)^{-2/3},
\end{equation}
equation (51) reduces to:
\begin{equation}
 kT = \frac{\mu m_p}{2\tilde{\beta}_{spec}}(\frac{1+\nu}{1-\nu})(\frac{2\pi G M}{t_0})^{2/3}[(H_0t_0)^{2/3}y].
\end{equation}
  Comparing this result with equation (25), we see that
\begin{equation}
  y = \frac{B}{A(Ht)^{2/3}}, 
\end{equation}
  {\it only if the energy is conserved} [i.e., assuming
  $E_f$ in equation (49) is equal to $E_i$].
On the other hand, by combining
equation (52) with equations (43), (44) and (47) and the virial theorem (14),
$y$ can be written as a function of $c$, for
a fixed density profile $f$:
\begin{equation}
 y(c,f) = \frac{\Delta_c^{1/3}(1-\nu)c \int_0^c f(x) g(x) x dx}{3\pi^{2/3} g^2(c)},
\end{equation}
 where we have assumed the boundary of the virialized region to be the radius at which the average density is
 $\Delta_c$ times the critical density of the universe (which
is usually chosen to be 200),
and $\nu$ is a function of $c$ and $f$ in equation (47).
Equation (55) fixes the concentration
parameter $c$ as a function of $y$
for a fixed density profile $f$,
which in turn is determined by equation (54).

The concentration parameter is fixed by the
     cosmology ($y$ parameter) as
   shown in Fig.(5). This relation can be well fit by:
\begin{equation}
   \log_{10}c = -0.17+ 1.2 \, y,
\end{equation}
for NFW profile, accurate to $5 \%$ in the range $0 < y < 1$.

 Let us now consider the evolution of $c$. We know that in an
 expanding universe $\Omega_m$ decreases with time.
 Comparing with Fig.(4) we see that in a flat $\Lambda$CDM
 universe $y$ is decreasing with time, while
 in an open/Einstein-de Sitter universe, it is almost constant.
  Then equation (56) implies that $c$ is a decreasing
 function of time (increasing function of redshift) in a $\Lambda$CDM
 universe, while it does not significantly evolve in an OCDM universe.
 As an example, the concentration parameter in an Einstein-de Sitter universe is about $40\%$ larger than that of a flat $\Lambda$CDM universe with $\Omega_m = 0.25$. This is consistent with the NFW results who find an increase of about $35 \%$ for $c$ as a function of mass in units of non-linear mass scale.

\clearpage
\myputfigure{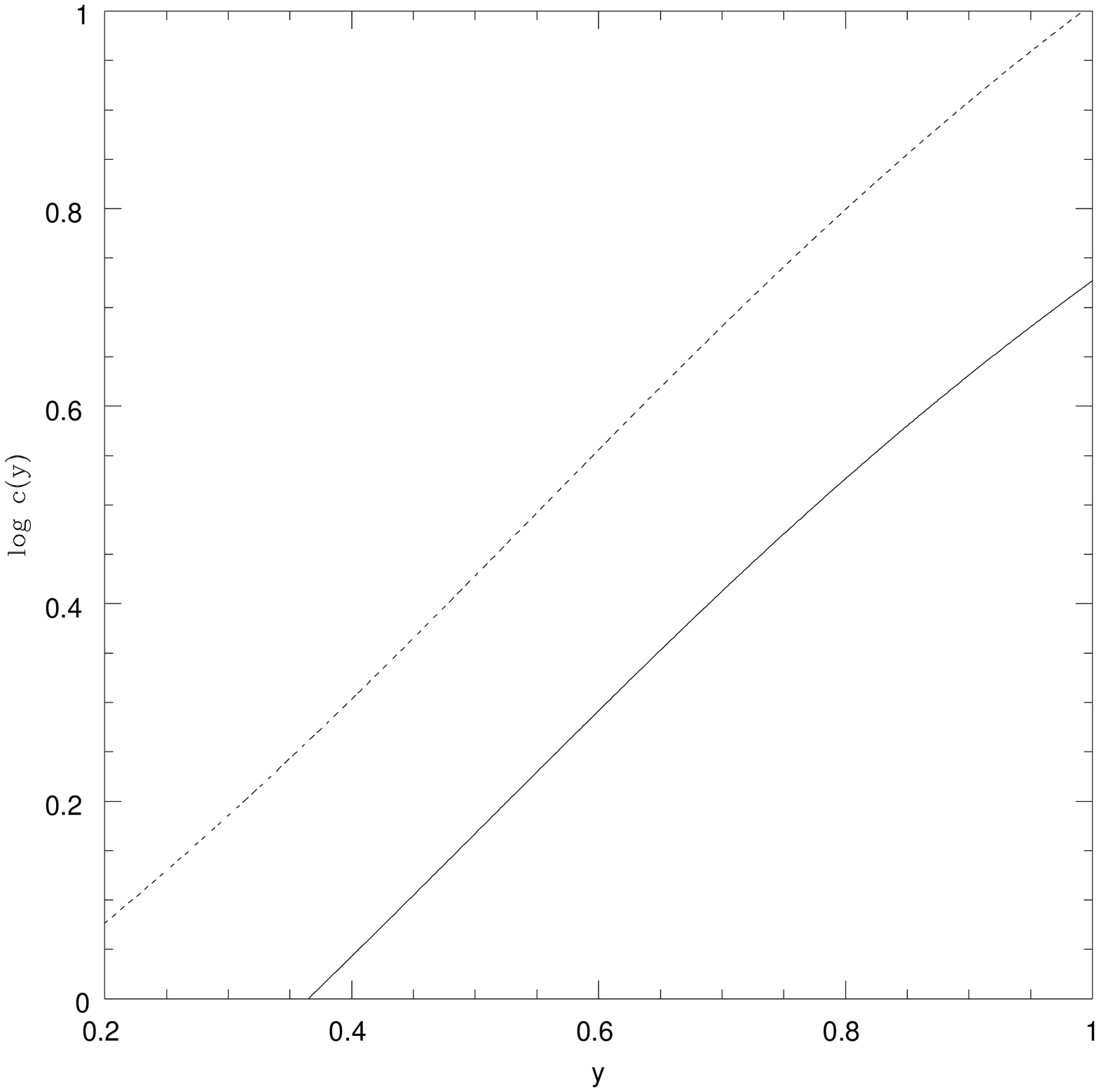}{6.0}{0.80}{-10}{-10} \vspace{-1.0cm}
\figcaption{Concentration parameter vs. $y$. The solid and dashed
curves are for Moore and NFW profiles, respectively.
\label{fig3c}} \vspace{\baselineskip}
\clearpage

 \myputfigure{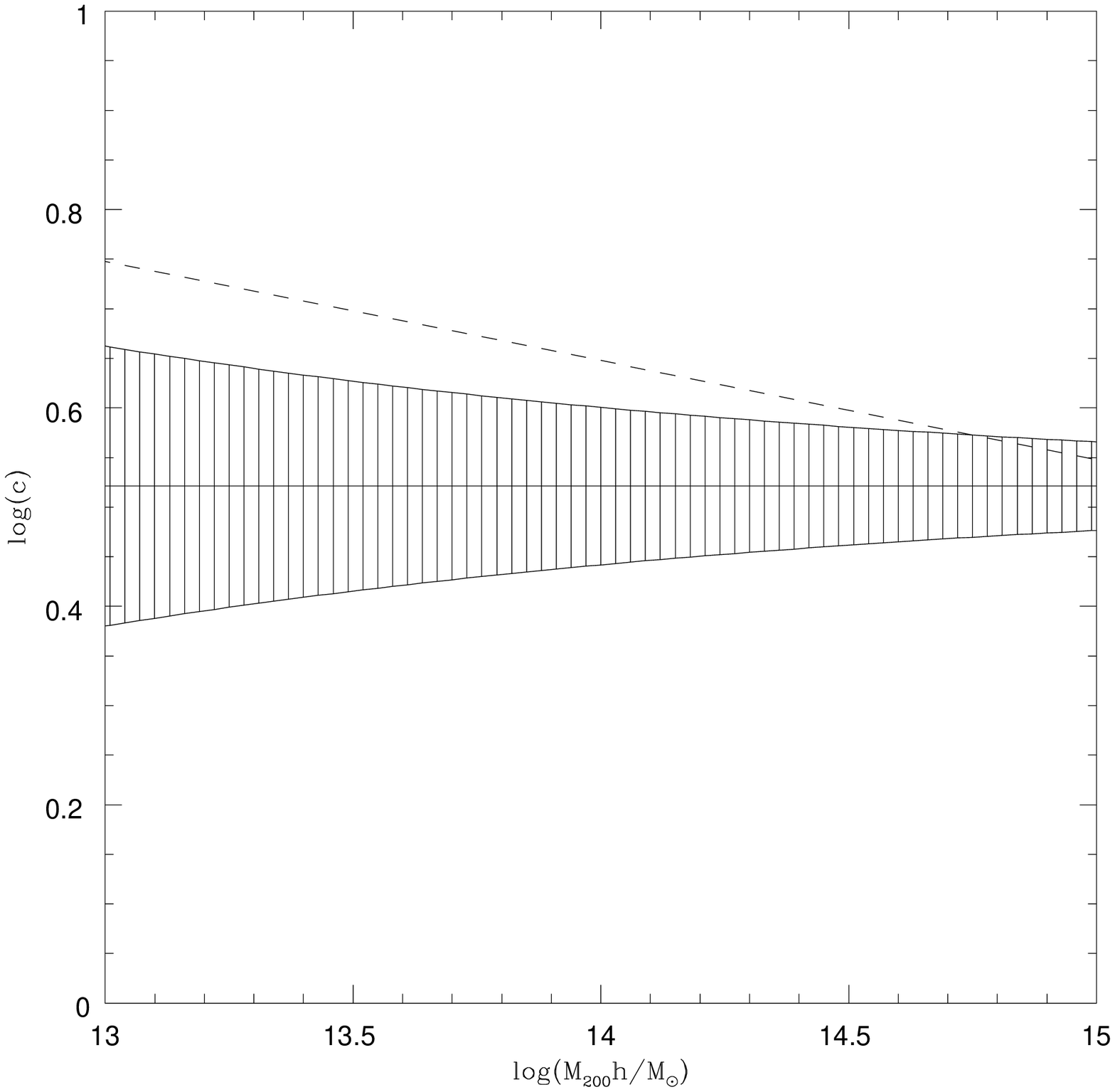}{6.0}{0.80}{-10}{-10}
\vspace{-1.0cm} \figcaption{ Predicted concentration parameter
($n=-1.5$ vs. mass for Thomas et al. 2001 simulation of an
Einstein-de Sitter cosmology. The shaded area indicates the $68\%$
likelihood region. The dashed line is the best fit to their
result. \label{fig7}} \vspace{\baselineskip}
\clearpage

 NFW simulations show a weak dependence of the concentration parameter on mass
 $ c \propto M^{-0.1}$. We see that our concentration parameter does not depend on mass.
 However its scatter is larger for small masses and so is
 marginally consistent with the simulation results (see Fig 6).
 Assuming that this discrepancy is only a consequence of non-spherical shape
 of original proto-cluster,
 in the next section, we attempt to modify the value of $y$ to match
 the simulation results.

\subsection{Corrections for Initial Non-Sphericity}

   In this section we try to incorporate the effects due to
 the non-spherical shape of the initial proto-cluster into our formalism.
 Unlike previous sections, the calculations of this section
 are
 not very rigorous and should be considered as an estimate of
 the actual corrections. In particular, these approximations
 lose accuracy if there are large deviations from sphericity
 which, as we see, is the case for low mass end of the M-T diagram.

  We are going to assume that non-sphericity comes in through
 a modifying factor $1+\cal{N}$ that only depends on the initial geometry
 of the collapsing domain,
 \begin{equation}
  y_{\cal{N}} = y (1+\cal{N}),
 \end{equation}
 where $y_{\cal{N}}$ is the modified value of $y$.
Next, let us assume that $R_i(\theta, \varphi)$ is the distance of
the surface of our collapsing domain from its center. We can
expand its deviation from the average in terms of spherical
harmonics $Y_{lm}(\theta,\varphi)$,
\begin{equation}
  \delta R_i(\theta, \varphi) = \sum_{l,m} a_{lm}
  Y_{lm}(\theta,\varphi).
\end{equation}
 If we try to write down a perturbative expansion for $\cal{N}$, the lowest order
 terms will be quadratic, since there is no rotationally invariant first order
 term. Moreover, having in mind that the gravitational dynamics is
 dominated by the large scale structure of the object, as an
 approximation, we are going to keep the lowest $l$ value. Since
 $l=1$ is only a translation of the sphere and does not change its
 geometry, the lowest non-vanishing multipoles are for $l=2$, and
 the only rotation invariant expression is
 \begin{equation}
  {\cal N} \approx \sum^{2}_{m=-2} |a_{2m}|^2,
 \end{equation}
 where we absorbed any constant factor in the definition of
 $a_{2m}$'s. The next simplifying assumption is that $a_{2m}$'s are
 Gaussian variables, with amplitudes proportional to the amplitude
 of the density fluctuations at the cluster mass scale. Mathematically, this is
 motivated by the fact that the concentration parameter predicted
 by simulations is closer to our prediction for spherical
 proto-clusters, at large mass end, where amplitude is smaller.
 The physical motivation is that since the density fluctuations
 decrease with scale, more massive clusters tend to deviate
 less from sphericity.
 Choosing Gaussian
 statistics for $a_{2m}$'s is only a simplifying assumption to carry out the
 calculations. Then it is easy to see that
\begin{equation}
 \Delta {\cal N}^2 = \frac{6}{25}<{\cal N}>^2,
\end{equation}
 and then using the definition of $\cal{N}$, assuming that it is a small
 correction we get
\begin{equation}
(\frac{\Delta y_{\cal N}}{y_{\cal N}})^2 \approx
\frac{6}{25}(1-\frac{y}{y_{\cal N}})^2+(\frac{\Delta y}{y})^2.
\end{equation}
  Note from equation (54), $\Delta y =
  (Ht)^{-2/3}(\Delta B/A)$ and $\Delta B/A$ is given in equation
  (37). We have also assumed that $\cal{N}$ and $y$ are statistically
  independent variables.
 In the next step, we define the amplitude of $\cal{N}$
\begin{equation}
  <{\cal N}> \approx \omega (\frac{\Delta B}{A})^2, \omega \approx 64
\end{equation}
 The numerical value of $\omega$ is fixed by plugging $y_{\cal{N}}$ into
 equation (56) to get the
 modified concentration parameter and comparing this result with the
 simulations of Thomas et al. (Fig 6). Fig 7 shows the
 modified concentration parameter as a function of mass
 in an Einstein-de Sitter universe. We see that the introduction
 of non-sphericity
 results in the cluster concentration parameter being
 a decreasing function of cluster mass with a scatter that
 also decreases with mass, in accord with simulations.
  Fig 8 shows the same comparison for a $\Lambda$CDM
 cosmology with Eke et al. (2001) fitting formula.
 We see that, although $\omega$ was obtained by fitting the Einstein-de Sitter
 simulations, our prediction is marginally consistent with the
$\Lambda$CDM
 simulations as well.\footnote{Eke et al. (2001) fitting formula was made
 for a lower mass range and its systematic error at the cluster
 mass range is indeed comparable to its difference with our
 prediction.}

\clearpage
 \myputfigure{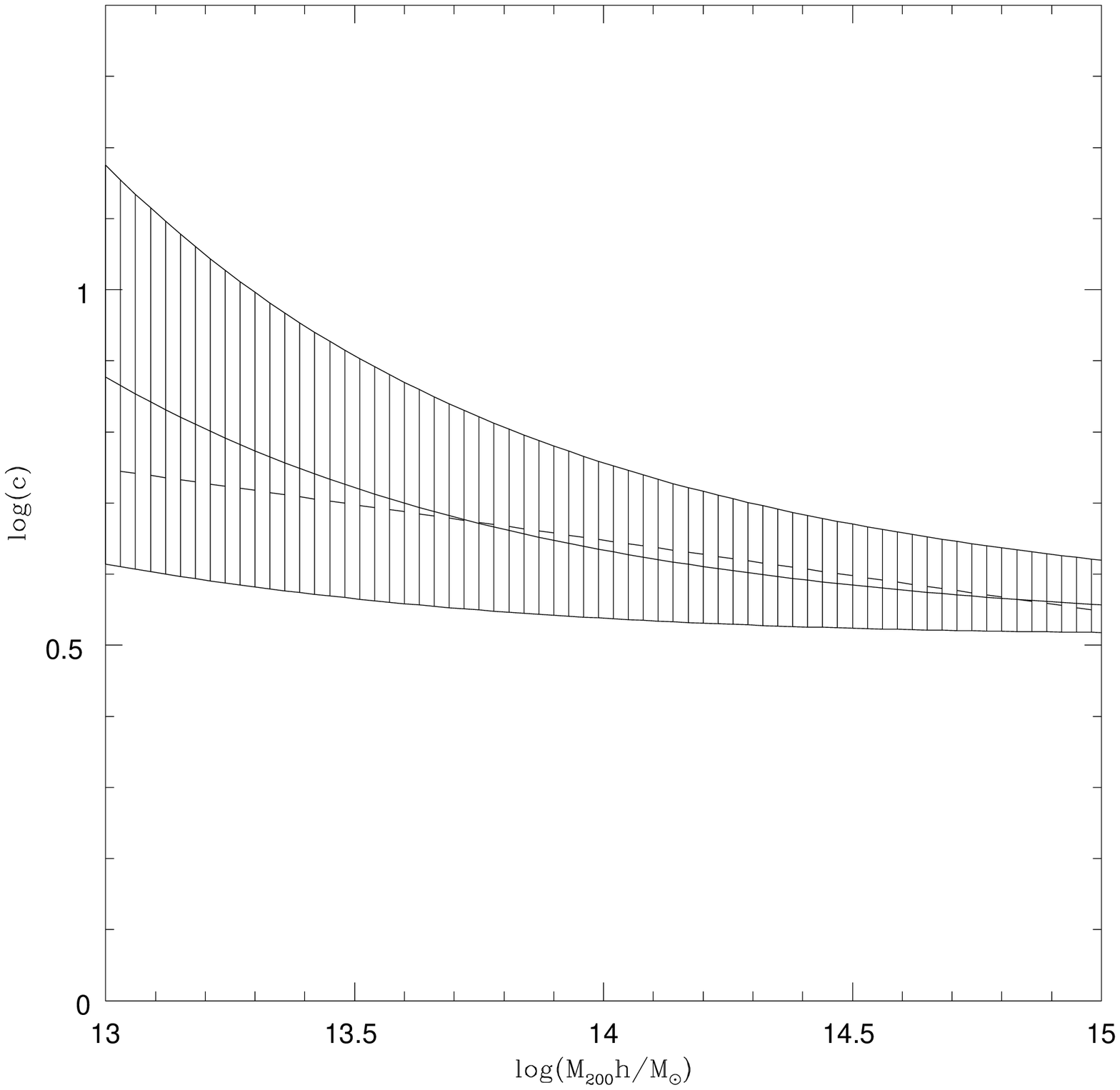}{6.0}{0.80}{-10}{-10}
\vspace{-1.0cm} \figcaption{ Similar to Fig 6 but with
non-spherical correction. The solid curve is fitted to match the
simulation result. \label{fig8}} \vspace{\baselineskip}
\clearpage 

 \myputfigure{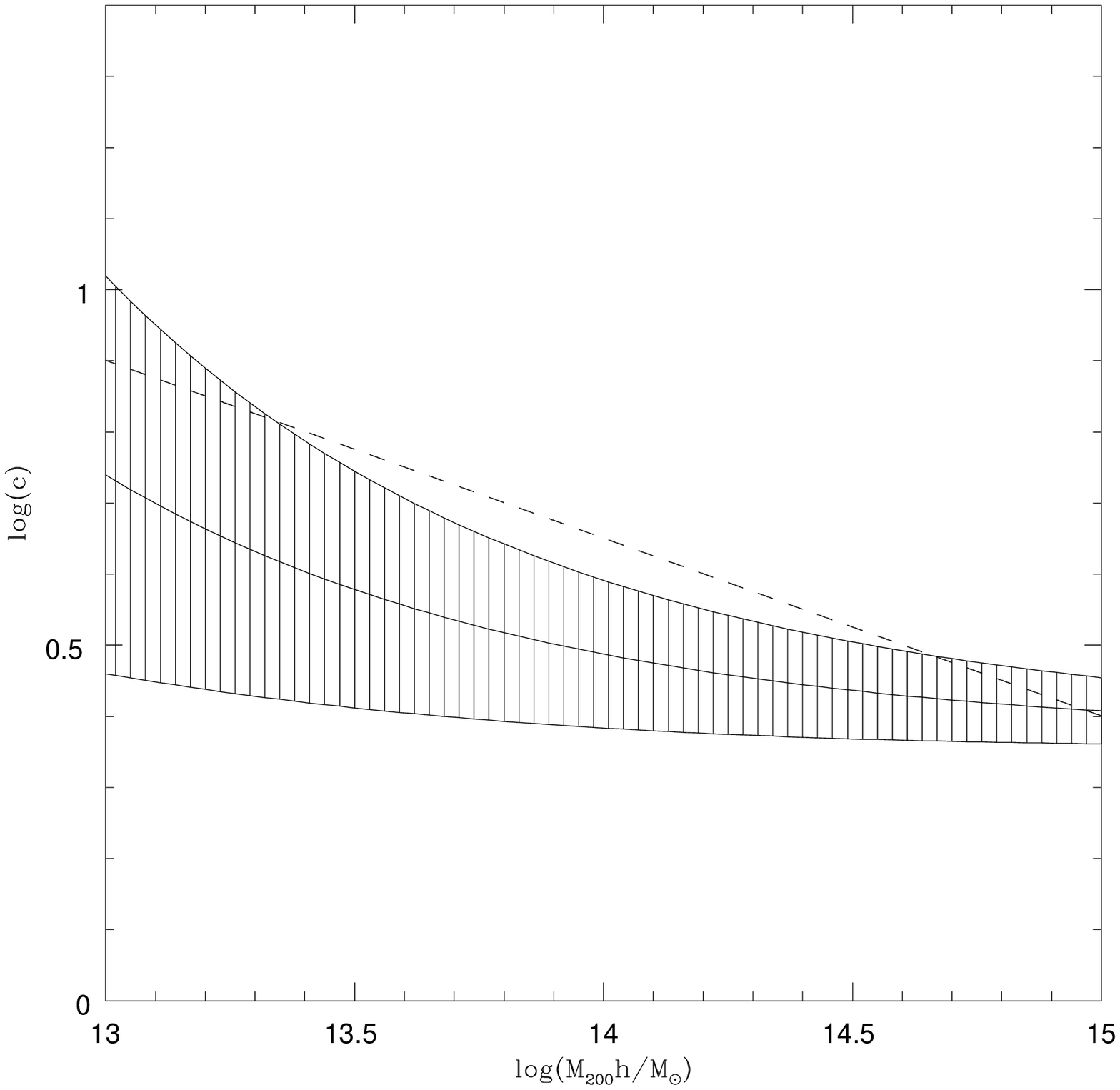}{6.0}{0.80}{-10}{-10}
\vspace{-1.0cm} \figcaption{The modified concentration parameter
vs. mass for a $\Lambda$CDM cosmology with $\Omega_m = 0.3$,
$\Omega_{\Lambda}=0.7$, $n=-1.5$ and $\sigma_8 = 0.9$. The dashed
line is the Eke
et al. (2001) fitting formula with $C_{\sigma} =23$ and the shaded area is 
the $68\%$
likelihood region. \label{figext}} \vspace{\baselineskip}
\clearpage

  \subsection{Scatter in M-T relation}
Since $y$ is linear in $B/A$, it also has a Gaussian
probability distribution function (PDF):
\begin{equation}
   P(y)dy = \frac{dy}{\sqrt{2\pi \Delta y^2}}
 \exp[-\frac{(y-\bar{y})^2}{2\Delta y^2}],
\end{equation}
    where $\bar{y}$ and $\Delta y$ are related to equations (36) and (37), by
 equation (54).
 As we mentioned above, the variation of $Q$ for the average
values of $y$ is negligible. However, the scatter in the value of
$y$ can be large, especially in the low mass end
(see equation 37)
and so the scatter
in $Q$ might become significant. In what follows we only consider
the NFW profile, since it is extensively considered in the
literature.
We find that the behavior of $Q$ can be fitted by:
\begin{equation}
  Q(y)^2 = Q^2_0+N(y-y_0)^2,
\end{equation}
where
\begin{equation}
Q_0 = 1.11,\, N = 1.8,\, y_0 = 0.538.
\end{equation}
The error in this fitting formula is less than $3\%$ in the range
$-1<y<2$. Inserting this into equation (63) leads to the PDF of $Q$:
\begin{eqnarray}
P(Q)dQ &= \frac{2 Q dQ}{\sqrt{2\pi \Delta y^2 N (Q^2 -Q^2_0)}} \nonumber \\
&\exp[-\frac{(y_0-\bar{y})^2 +(Q^2-Q^2_0)/N}{2\Delta
y^2}]\cosh[\frac{y_0-\bar{y}}{\Delta y^2}
\sqrt{\frac{Q^2-Q^2_0}{N}}].
\end{eqnarray}

Fig 9 shows three different examples of the PDF obtained here. It
is clear that the scatter in $Q$ is asymmetric. In fact, since the
average value of $y$ is close to the minimum of equation (57), the
scatter in $y$ shifts the average value of $Q$ upwards
systematically. In the limit of large $\Delta y$, where this shift
is significant, $P(Q)$ is approximately:
\begin{equation}
 P(Q)dQ = \frac{2 Q dQ}{\sqrt{2\pi \Delta y^2 N (Q^2
-Q^2_0)}}\exp[-\frac{(Q^2-Q^2_0)}{2N\Delta y^2}]\, \Theta(Q-Q_0),
\end{equation}
 where $\Theta$ is the Heaviside step function.

\clearpage
\myputfigure{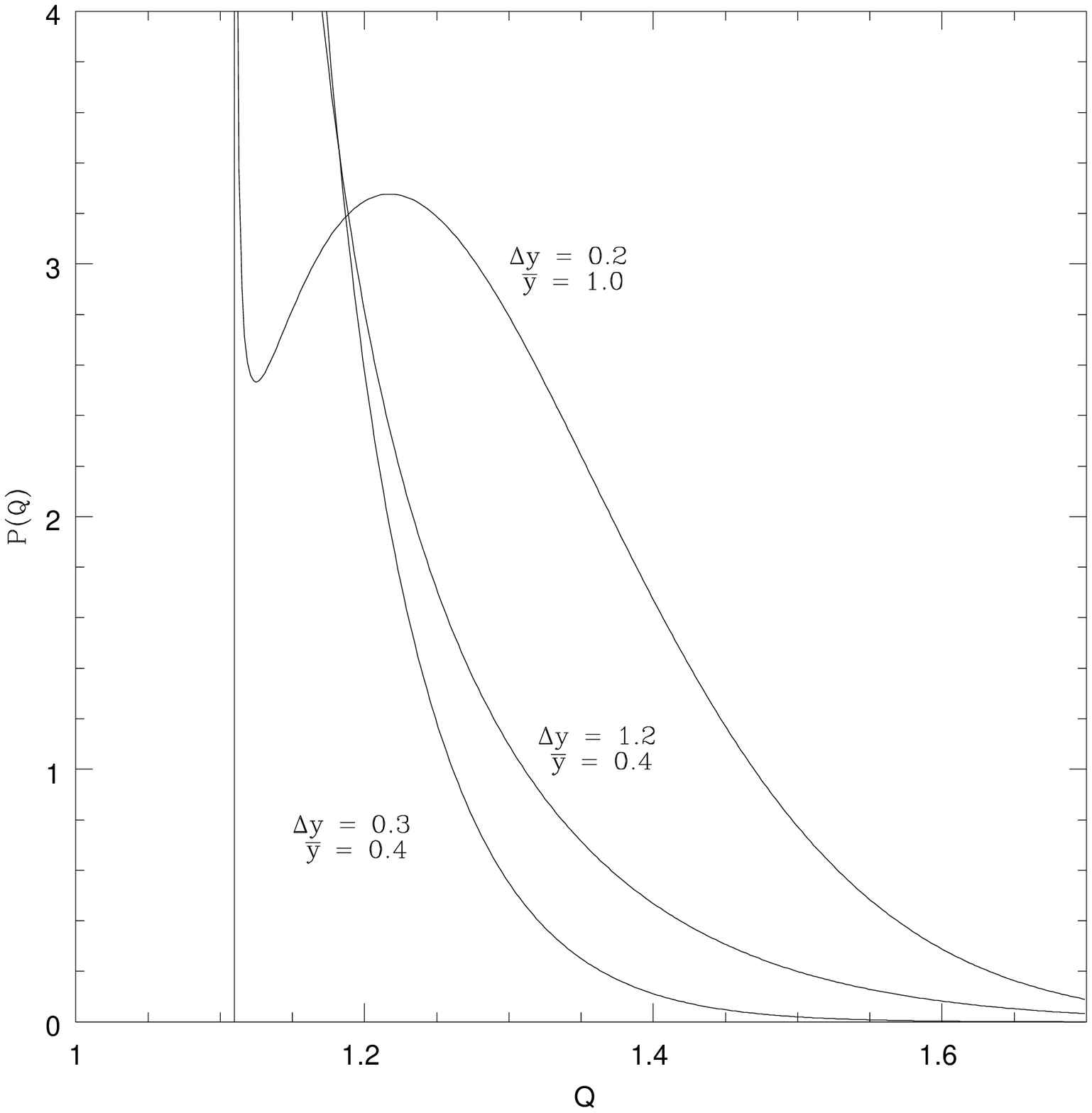}{6.0}{0.80}{-10}{-10} \vspace{-1.0cm}
\figcaption{Probability distribution function of normalization for
three different $\bar{y}$ and $\Delta y$'s. The PDF vanishes for
$Q<Q_0 = 1.11$. The large value of $\bar{y} = 1.0$ is possible if
one includes non-spherical corrections (\S 3.6). \label{fig4}}
\vspace{\baselineskip}
\clearpage

 Although the assumption of Gaussianity is not strictly valid for
 $y_{\cal{N}}$ which includes non-spherical corrections, we still can,
 as an approximation, use the above expressions for the PDF by replacing
 $\bar{y}$ and $\Delta{y}$ by $\bar{y}_{\cal{N}}$ and $\Delta y_{\cal{N}}$.
  It is also easy to find the average of $Q^2$
$$
 <Q^2> = Q^2_0 + N[(\bar{y}-y_0)^2+\Delta y^2].
$$
  As an example, for $\bar{y} \sim y_0$, while $\Delta y = 0.5$
gives only about $16 \%$ systematic increase in $Q$,  $\Delta y = 1.0$ leads to$\sim 60 \%$ increase.

\subsection{Observed and Average Temperatures}
   The temperature found in \S 2.4 is the density weighted
   temperature of the cluster averaged over the entire
   cluster.
 However, the observed temperature, $T_f$, can be considered as
a flux-weighted spectral temperature averaged over a smaller,
central region of the cluster. The two temperatures may be
different, due to presence of inhomogeneities in temperature. We
use the simulation results by Mathiesen \& Evrard (2001) to relate
these two temperatures and refer the reader to their paper for
their exact definitions and the discussion of the effects which
lead to this difference:
\begin{equation}
 T = T_f [ 1+ (0.22\pm 0.05)\log_{10} T_f(\keV) - (0.11\pm 0.03)].
\end{equation}
 This correction changes the mass-temperature relation from $M \propto T^{1.5}$ to $M \propto T^{1.64}$ for
the observed X-ray temperatures. We use this correction in
converting the observed X-ray temperature to virial temperature in
Figures 1 and 10-12.

\section{Predictions vs. Observations}

 \subsection{Power Index}
It is clear from equation (49) that we arrive at the usual $M
\propto T^{1.5}$ relation that is expected
  from simple scaling arguments and is consistent with the numerical simulations (e.g. EMN or Bryan \& Norman 1998). 
 On the other hand, the observational $\beta$-model mass estimates
 lead to a steeper power index in the
 range $1.7-1.8$. Although originally interpreted as an artifact of the
 $\beta$-model (HMS), the same behavior was seen for masses estimated from
 resolved temperature profile (FRB). FRB carefully analyzed the data and
 interpret this behavior as a bent in the M-T at low temperatures. This
was
 confirmed by
 Xu, Jin \& Wu (2001) who found the break at $T_X = 3-4\,\keV$.

As discussed in \S 3.7, the asymmetric scatter in $Q$ introduces a
systematic shift
 in the M-T relation.
 For large values of $\Delta y$ all of the temperatures are larger
 than the value given by the scaling relation (54) for
 average value of $y$.
This scatter increases for smaller masses (see equation 37),
 hence smaller clusters are hotter than the scaling prediction.
 As a result, the M-T relation becomes steeper
 in the low mass range as $Q$ increases, while the intrinsic scatter of the data is also
 getting larger.
 Indeed, increased scatter is also observed in the FRB data(Fig 1 and Fig's
 10-12
 to compare with our prediction) but they
interpret it as the effect of different formation redshifts. We
will address this interpretation in \S 5.

\subsection{Normalization}

     As we discussed in \S 3.4, our normalization
     (i.e., $Q$ in equation 50)
     is rather stable with
respect to variations of cosmology and the equilibrium density profile.
Table 1 compares this value with various
 observational and simulation results.
 \linebreak
 \linebreak

\clearpage
\begin{tabular}{|c|c|c|}
 \hline
$Q$  &Method & Reference  \\ \hline
$1.12 \pm 0.02$ & Analytic  & This paper  \\
$1.21 \pm 0.06$ & Hydro-Simulation & Thomas et al. 2001 \\
$1.05 \pm 0.13$ & Hydro-Simulation & Bryan \& Norman 1998 (BN) \\
$1.22 \pm 0.11$ & Hydro-Simulation & Evrard et al. 1996 (EMN)\\
$1.32 \pm 0.17$ & Optical mass estimate &  Horner et al. 1999 (HMS1)  \\
$1.70 \pm 0.20$ & Resolved temperature profile & Horner et al. 1999 (HMS2)\\
$1.70 \pm 0.18$ & Resolved temperature profile & Finoguenov et al.
2001 (FRB) \\ \hline
\end{tabular}

\begin{center}
Table 1. Comparison of normalizations from different methods. The
last two rows only include clusters hotter than 3 keV.
\end{center}
\vspace{5mm}
\clearpage

We see that our analytical method
 is consistent with
 the hydro-simulation results,
 indicating validity of our method,
 since both have a similar physics input and the value
 of $\beta_{spec}$ used here was
 obtained from simulations.
 On the other hand, X-ray mass estimates lead
 to normalizations about $50\%$ higher than our result and simulations. The
 result of the optical mass estimates quoted
 is marginally consistent with our result.
Assuming that this is true would imply that optical masses are
systematically higher than
 X-ray masses by $\sim 80\%$. Aaron \etal (1999)
 have compared optical and X-ray masses for
 a sample of 14 clusters and found, on the contrary,  a systematic difference less than $10\%$.
 The optical masses used in the work of Aaron \etal (1999)
 were all derived by CNOC group (Carlberg et al. 1996) while the Girardi
 et al. 1998 data, used in HMS analysis quoted above, was compiled from
different
 sources and has larger scatter and unknown systematic errors. In fact
HMS excluded
 a number of outliers to get the correct slope and original data had even larger scatter (see their Fig 1).
 Therefore it may be
 that the systematic error in the optical result of Table 1, be much larger and
 so in agreement with other observations.

As discussed in \S 3.1,  one possible source for difference
between theoretical and observational normalizations is that the
values for $\beta_{spec}$ are different in the two cases due to
systematic selection effects. Also, intriguingly,
   Bryan \& Norman (1998) show that there is
   a systematic increase in the obtained value of $\beta_{spec}$ by increasing the
    resolution of the simulations.
    Whether this is a significant and/or real
    effect for even higher resolutions is not clear to us.

 However, the fact that the slope is unchanged indicates
 that the missing process is, probably,  happening at small
 scales and so relates the intermediate-scale temperature to the small-scale flux-weighted spectral temperature
 by a constant factor, independent of the large-scale structure of the cluster. In this case the actual value
 of $\beta_{spec}$ must be $\sim 0.6$.

 Figures 10-12 show the prediction of our model, shifted
 downwards to fit the observational data in the massive end,
 versus the observational data
 of FRB using resolved temperature profile and corrected as discussed in
 \S 3.8. The correction due to initial non-sphericity (\S 3.6) is included
 in the theoretical plot. The value of $\sigma_8$ which enters
 $\Delta B/A$ (equation 37) through $M_{0L}$, is fixed
 by cluster abundance observations (e.g. Bahcal \& Fan 1998).
\pagebreak

\clearpage
\myputfigure{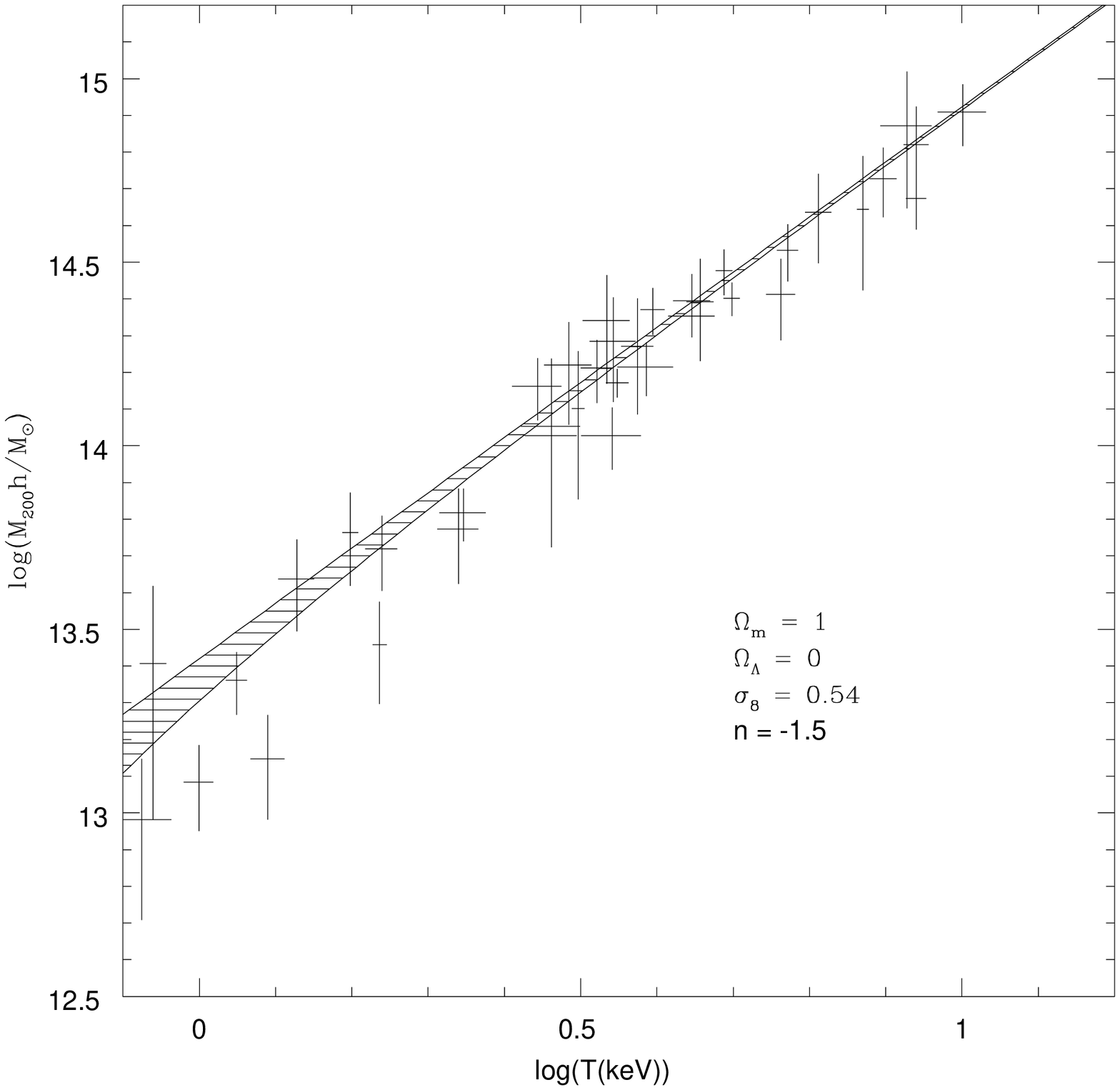}{6.0}{0.80}{-10}{-10} \vspace{-1.0cm}
\figcaption{ Our result with shifted normalization for an
Einstein-de Sitter universe vs. FRB data. The shaded area
indicates the $68 \%$ confidence level region (\S 3.7).
\label{fig5a}} \vspace{\baselineskip}
\clearpage

\myputfigure{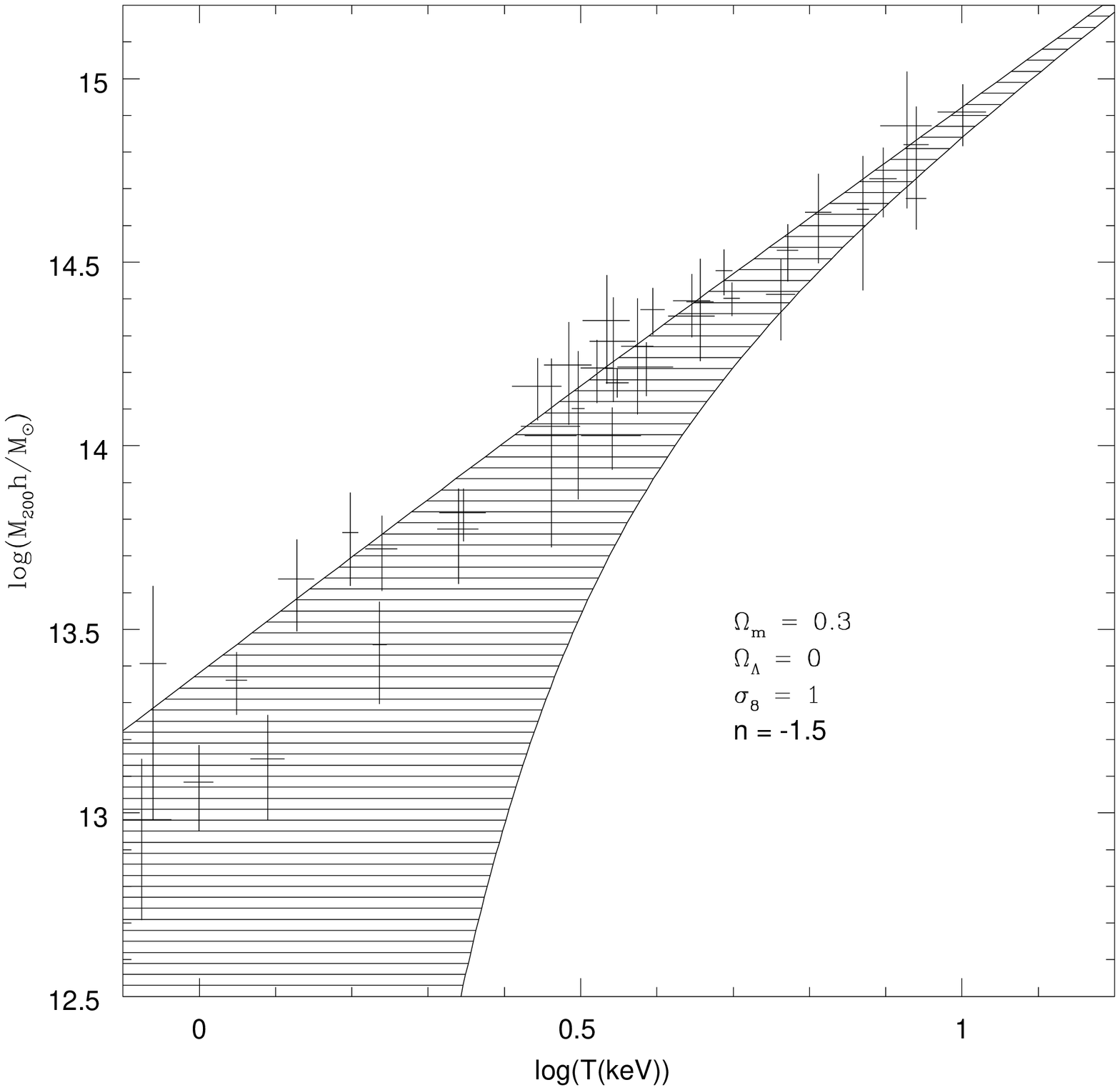}{6.0}{0.80}{-10}{-10} \vspace{-1.0cm}
\figcaption{ Our result with shifted normalization for an OCDM
universe vs. FRB data.  The shaded area indicates the $68 \%$
confidence level region (\S 3.7).\label{fig5b}}
\vspace{\baselineskip}
\clearpage

\myputfigure{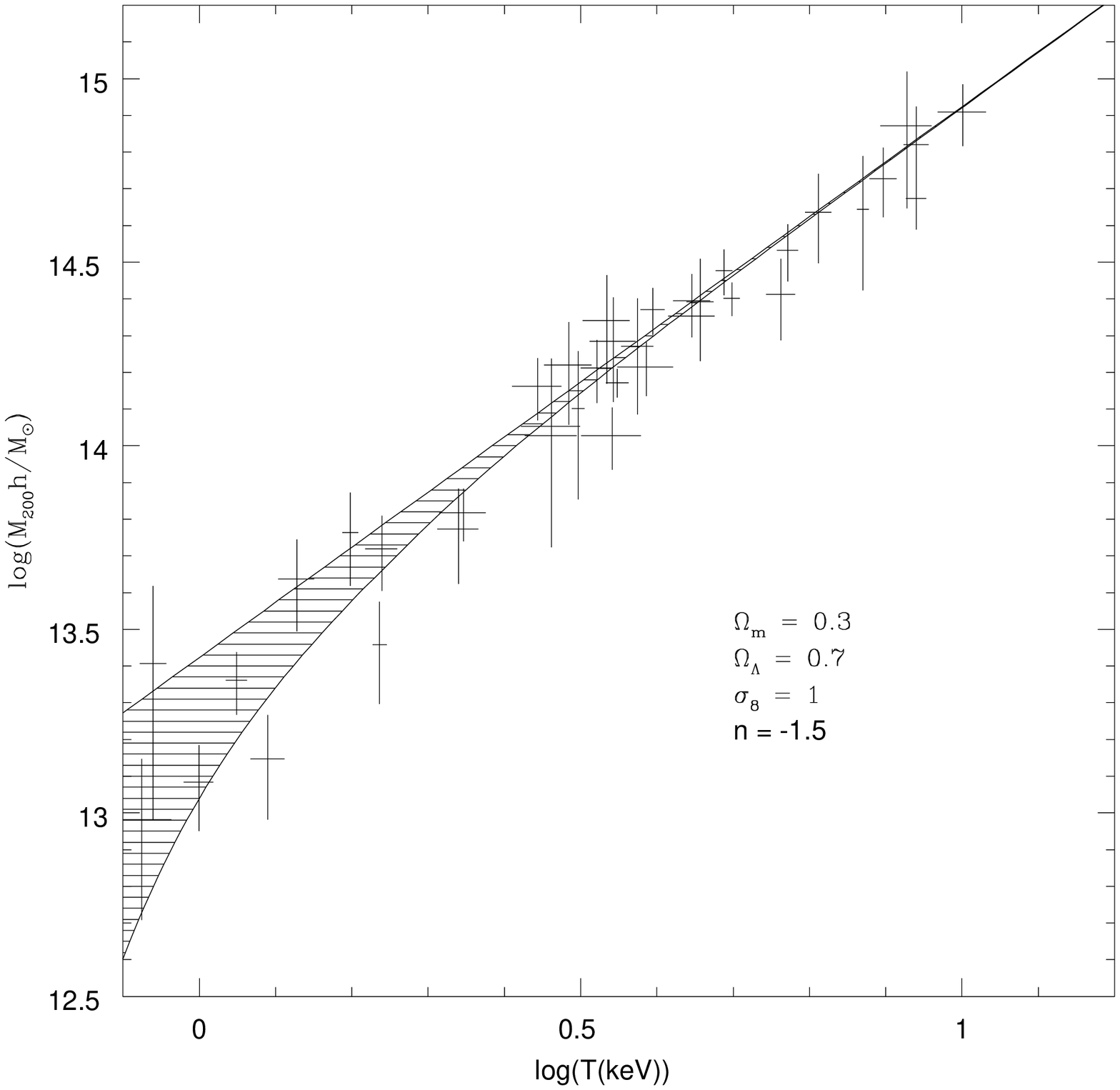}{6.0}{0.80}{-10}{-10} \vspace{-1.0cm}
\figcaption{ Our result with shifted normalization for a
$\Lambda$CDM universe vs. FRB data. The shaded area indicates the
$68 \%$ confidence level region (\S 3.7). \label{fig5c}}
\vspace{\baselineskip}
\clearpage

We see that while an Einstein-de Sitter cosmology under-estimates
the scatter in the low mass end, a typical low density OCDM cosmology
overestimates it. On the other hand, a typical $\Lambda$CDM
cosmology is consistent with the observed scatter. Interestingly,
this is consistent with various other methods, in particular
CMB+SNe Ia
result which point to a low-density flat universe 
(de Bernardis \etal 2000; Balbi \etal 2000; Riess \etal 1998).

  \subsection{Evolution of M-T Relation}
 As we discussed in \S 3.4, the value of our normalization has a weak dependence on cosmology.
 Going back to equation (49), we see that the time dependence of the M-T
relation is simply:
 $M \propto H^{-1} T^{1.5}$. Assuming a constant value of $\beta_{spec}$, this formula can be potentially used
 to measure the value of $H$ at high redshifts and so constrain the
 cosmology.

  Schindler (1999) has compiled a sample of 11 high redshift
  clusters ($0.3 < z < 1.1 $) from the literature with measured
  isothermal $\beta$-model masses. In these estimates, the gas is
  assumed to be isothermal and have the density profile:
  \begin{equation}
   \rho_g(r)= \rho_g(0) (1+(\frac{r}{r_c})^2)^{-3\beta_{fit}/2}
  \end{equation}
  Then, the mass in overdensity $\Delta_c$ is given by:
  \begin{equation}
  M \simeq (\frac{3 H^2
  \Delta_c}{2 G})^{-1/2}(\frac{3 \beta_{fit} k T}{G \mu m_p})^{3/2}.
  \end{equation}

\noindent
Comparing this with equation (49), and neglecting the difference between
virial and X-ray temperatures,
  we get:
  \begin{equation}
  Q \simeq (3 \pi)^{-2/3}(\tilde{\beta}_{spec}/\beta_{fit})
  \Delta_c^{1/3}
  \end{equation}
   In the last two equations, we have ignored $r_c$ with respect
   to the radius of the virialized region, which introduces less
   than $3 \%$ error. 
   This allows us to find the normalization $Q$ from
   the value of $\beta_{fit}$ (independent of cosmology in this case).

   Assuming $\tilde{\beta}_{spec} = 0.9$, equation (71) gives the value of
   $Q$ for a given  $\beta_{fit}$. Fig 13 shows the value of $Q$
   versus redshift for Schindler (1999) sample and also FRB resolved temperature
   method for low redshift clusters. This result is consistent
   with no redshift dependence and the best fit is:
   \begin{equation}
   \log Q = 0.23 \pm 0.01 (systematic) \pm 0.04 (random) + (0.09  \pm 0.04)z.
   \end{equation}

\clearpage
\myputfigure{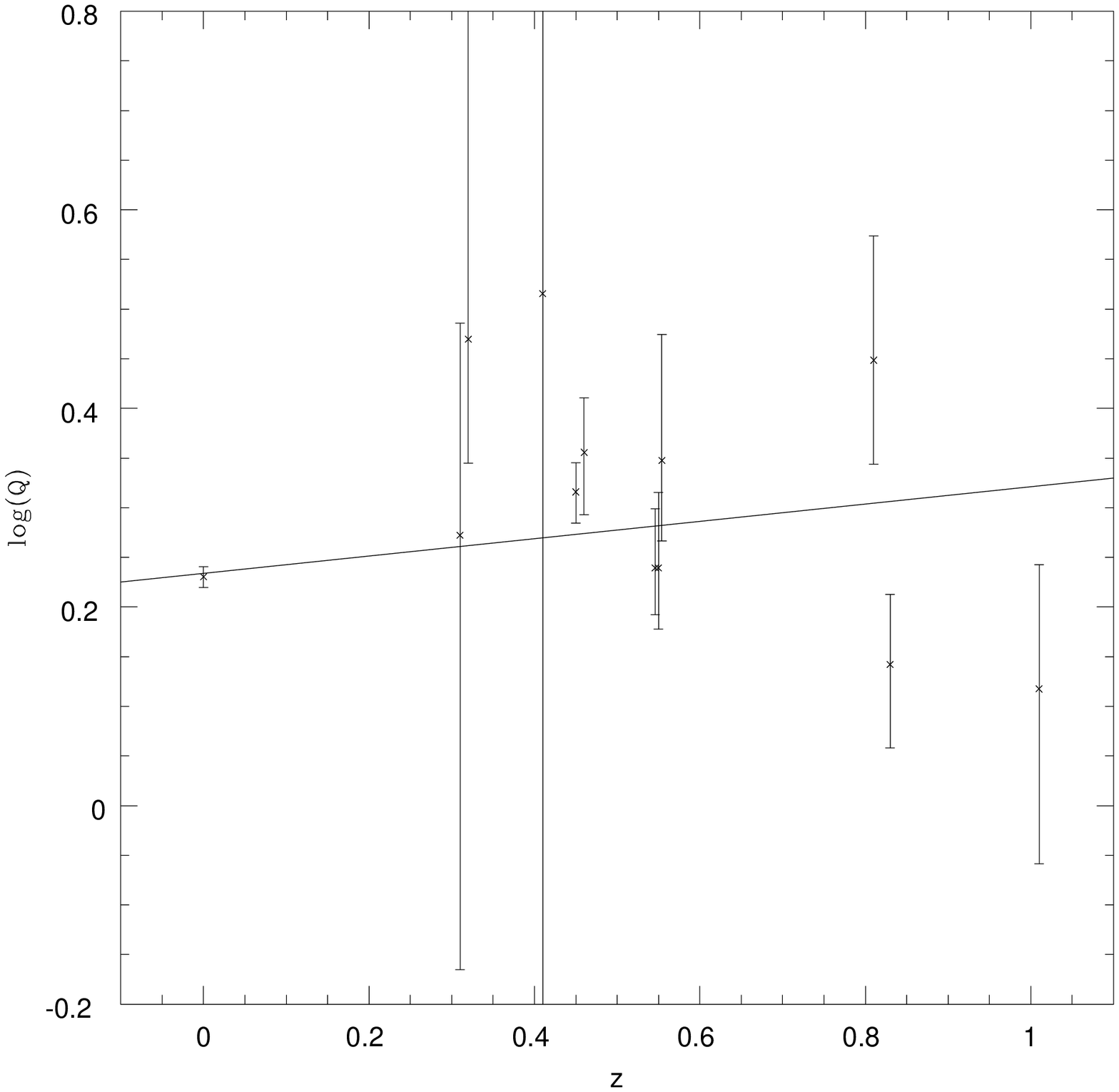}{6.0}{0.80}{-10}{-10} \vspace{-1.0cm}
\figcaption{ Evolution of the normalization $Q$ with redshift. The
data points are from Schindler (1999) high-z sample and FRB low-z
normalization at $z = 0$. The line is the best fit. \label{fig6}}
\vspace{\baselineskip}
\clearpage

\noindent The combination of this result and equation (50) gives
    \begin{equation}
       kT = (11.2 \pm 1.1 \keV) e^{(0.21\pm 0.09)z}(\frac{M}{10^{15} h^{-1}
       M_{\odot}})^{2/3}
    \end{equation}

\noindent  which relates  X-ray temperature of galaxy clusters to their
    masses that does not depend on the
    theoretical uncertainties with regard
    to the normalization coefficient of the
    M-T relation, in this range of redshifts.
    The systematic error in this result is less than $5\%$ while
    the random scatter can be as large as $20\%$. This result is
    valid for $ T_X > 4 \keV$ since below this temperature the
    systematic shift due to random scatter becomes significant (\S 3.7).
    It is easy
    to see that
    this threshold moves to lower temperatures in high redshifts
    in our formalism.

     Note that the more realistic interpretation of a possible evolution
  in observed $Q$, obtained above, is that in equation (71), $Q$ remains
constant (Fig 5) and, instead,
  $\tilde{\beta}_{spec}$ varies with time.
 However there will be virtually no difference
 with respect to the M-T relation and, moreover, the weak redshift dependence
 in equation (73) shows that  $\tilde{\beta}_{spec}$ is indeed almost constant.

\section{Discussion}

  In this section we discuss the validity of different approximations which were adopted throughout this paper.

  In the calculation of initial and final energy of the cluster, we ignored any contribution from the
 vacuum energy. In fact we know that cosmological constant in the Newtonian limit can be considered
 as ordinary matter with a constant density. If the cosmological constant does not change with time,
 then its effect can be considered as a conservative force and so energy is conserved. However, in both
 initial state and final equilibrium state of the cluster the density of cosmological constant is much
 smaller than the density of matter and hence its contribution is negligible. This does not hold for
 quintessential models of the vacuum energy since $\Lambda$ changes
 with time and energy is not conserved.
 This may be used as a potential method to distinguish
 these models from a simple cosmological constant.

 Let us make a simple estimate of the importance of this effect.
 We expect the relative contribution of a varying cosmological
 constant be maximum when the density of the proto-cluster is minimum.
 This happens at the turn-over radius  which is almost twice $r_v$, the
 virial radius, in the top-hat approximation. Then the contribution
 to the energy due to the vacuum energy would be:
\begin{equation}
 \frac{\delta E}{E} \sim (\frac{4 \pi}{3} \Lambda (t_0/2) (2 r_v)^3)/M
= (\frac{8}{200})(\frac{\Lambda (t_0/2)}{\rho_c(t_0)}) =
(\frac{8}{200})(\frac{H^2(t_0/2)}{H^2(t_0)}) \Omega_{\Lambda}(t_0/2) \sim 0.1.
\end{equation}
\noindent
This gives about $10 \%$ correction to $y$ or about $15 \%$ correction
to $c$. We see that still the effect is small but, in principle observable
if we have a large sample of clusters with measured concentration parameters.

 A systematic error in our results might have been introduced by
 replacing the density distribution by an averaged radial
 profile which under-estimates the magnitude of gravitational energy
 and so the temperature. Also, as shown by Thomas et al. (2000),
 most of the clusters in simulations are either steeper or
 shallower than the NFW profile at large radii. Since our normalization
 is consistent with simulation results, we think that these
 effects do not significantly alter our predictions.

 Let us now compare our results with that of Voit (2000),
 who made the first such analytic calculation (but we were not aware
 of his elegant work until the current work was near completion)
 which
is not
    based on the top-hat initial density perturbation and uses
 the same ingredients to obtain M-T relation. First of all, as noted
 in \S 3.3, his result, which is equivalent to the central peak
 approximation to find $B/A$, ignores the possibility of mergers.
 As we see in Fig (2) the value of $B/A$ and so $y$ is about $50\%$
 larger in the single peak
 case than the multiple peak case.
 Although it does not change the normalization very much, it
 overestimates the concentration parameter (Fig (5)) in our formalism.
 However, as pointed out by the referee, our formaism for finding the value
 of $c$ is based on the assumption of isotropic velocity dispersion
 (equation (45)) and hence may not be directly compared to Voit (2000)
 who does not make such assumption. Also Voit (2000) 
 has neglected the cosmology dependence of
 the surface correction, $\frac{1+\nu}{1-\nu}$, which gives a false cosmology
 dependence to the normalization. This is
 inconsistent with hydro-simulations (e.g. EMN, Mathiesen 2000) and
 does not give the systematic shift in the lower mass end, if we assume 
 it does not have a non-gravitational origin.

   Finally, we comment on the interpretation of the scatter/bent in the lower
   mass end as being merely due to different formation redshifts which is suggested
   by FRB. Although different formation redshifts can certainly produce this scatter/bent, it is not possible to distinguish it from scatter in initial
   energy of the cluster or its initial non-sphericity, in our formalism. On the
   other hand, the FRB prescription, which assumes constant temperature
   after the formation time, is not strictly true because of the on-going
   accretion of matter even after the cluster is formed. 
   So, as argued by Mathiesen 2000 using simulation results, the
   formation time might not be an important factor,
   whereas the effect
   can be produced by scatter in initial conditions of the proto-cluster.

\section{Conclusion}

    We combine conservation of energy with the virial theorem to
    derive the mass-temperature relation of the clusters of galaxies
    and obtain the following results:
    \begin{itemize}

    \item Simple spherical model gives the usual relation 
    $T=CM^{2/3}$, with the normalization factor $C$
    being in excellent agreement with hydro simulations.
    However, both our normalization and that from 
    hydro simulations are about $50 \%$ higher than the
    X-ray mass estimate results. 
    This is probably due to our poor understanding
    of the history of the cluster gas.

    \item  Non-sphericity introduces an asymmetric, mass-dependent scatter 
    (the lower the mass, the larger the scatter)
    for the $M-T$ relation thus alters the slope at the low mass end
    ($T \sim 3 \keV$).
    We can reproduce the recently observed scatter/bent in the M-T
    relation in the lower mass end for a low density
    $\Lambda$CDM cosmology, while Einstein- de Sitter/OCDM
    cosmologies under/overestimate this scatter/bent. We conclude
    that the behavior at the low mass end of the M-T digram can be
    used to constrain cosmological models.

    \item  
    We point out that the concentration parameter of the cluster
    and its scatter can be determined by our formalism. 
    The concentration parameter determined using this method 
    is marginally consistent with simulation results, 
    which provides a way to find non-spherical corrections to M-T
    relation by fitting our concentration parameter to the simulation
    results.

    \item Our normalization has a very weak dependence on
    cosmology and formation history. This is consistent with
    simulation results.

    \item We find mass-temperature relation (73), for clusters of
    galaxies,
    based on the observations calibrated by our formalism,
    which can be used to find masses of galaxy clusters
    from
    their X-ray temperature in the range of redshift $ 0< z < 1.1$
    with the accuracy of $20\%$. This is a powerful tool to
    find the evolution of mass function of clusters, using their temperature
    function.

    \end{itemize}

\acknowledgments This research is supported in part by grants
NAG5-8365. N.A. wishes to thank Ian dell'Antonio, Licia Verde
and Eiichiro Komatsu for useful discussions.

\appendix
\section{Statistics of $b/a$}

  $a$ and $b$ are defined as:
\begin{equation}
 a = \int_0^1\delta_i(\tilde{x}) d^3x,\, b = \int_0^1 (1-x^2) \delta_i(\tilde{x}) d^3x
\end{equation}
  Assuming a Gaussian statistics for the linear density field, The Probability
Distribution Function for $a$ and $b$, takes a Gaussian form:
\begin{eqnarray}
P(a,b)\, da\, db &=& \frac{1}{2\pi\sqrt{L}}\exp[-\frac{1}{2L}(<b^2>a^2+<a^2>b^2+2<ab>ab ) ]\, da\, db, \nonumber \\
 L &=& <a^2><b^2>-<ab>^2.
\end{eqnarray}

   Then, for fixed $a$, we have:
\begin{eqnarray}
\frac{<b>}{a} &=& \frac{<ab>}{<a^2>} \\
\Delta b &=& \sqrt{\frac{L}{<a^2>}}.
\end{eqnarray}
   To find the quadratic moments of $a$ and $b$, we assume a power-law
 linear correlation function:
\begin{equation}
 \xi_i({\mathbf r}) = <\delta_i({\mathbf x})\delta_i({\mathbf x + r})>
 = (\frac{r_{0i}}{r})^{3+n}.
\end{equation}
    The moments become:
\begin{eqnarray}
<a^2> &=& 8 \pi^2 (\frac{r_{0i}}{R_i})^{3+n} F_{00}(n), \\
<ab> &=& 8 \pi^2 (\frac{r_{0i}}{R_i})^{3+n} (F_{00}(n)-F_{02}(n)), \\
<b^2> &=& 8 \pi^2 (\frac{r_{0i}}{R_i})^{3+n}(F_{00}(n)-2F_{02}(n)+F_{22}(n)),
\end{eqnarray}

    where $F_{ml}(n)$ is defined as:
\begin{equation}
  F_{ml}(n) \equiv \frac{1}{8 \pi^2}\int x^m_1 x^l_2 |{\mathbf x_1 -x_2}|^{-(n+3)} d^3 x_1 d^3 x_2,
\end{equation}
    where the integral is taken inside the unit sphere. Taking the angular parts of the integral, this reduces to:
\begin{equation}
  F_{ml} = \frac{1}{n+1}\int_0^1 \int_0^1 dx_1 dx_2 x_1^{m+1} x_2^{l+1}
  (|x_1-x_2|^{-1-n}-|x_1+x_2|^{-1-n}).
\end{equation}
   Then, taking this integral for the relevant values of $m$ and $l$, and
   inserting
 the result into (A6-A8) and subsequently (A3-A4) gives:
\begin{equation}
   \frac{<b>}{a} = \frac{4(1-n)}{(n-5)(n-2)},
\end{equation}
 with
\begin{equation}
 \Delta{b} = \frac{16 \pi 2^{-n/2}}{(5-n)(2-n)}[\frac{n+3}{n(7-n)(n-3)}]^{\frac{1}{2}}(\frac{r_{0i}}{R_i})^{\frac{n+3}{2}}.
\end{equation}

\begin{references}
\reference{} Aaron, L.D., Ellingson, E., Morris, S.L., Carlberg,R.G. 1999, ApJ, 517, 2, 587 
\reference{} Bahcall, N.A.,
\& Cen, R. 1992, ApJ, 398, L81 
\reference{} Bahcall, N.A., Fan,
X., \& Cen, R. 1997, ApJ, 485, L53 
\reference{} Bahcall, N.A., \& Fan, X. 1998, ApJ, 504, 1 
\reference{} Balbi, A., \etal 2000, ApJ, 545, L1
\reference{} Bialek, J.J., Evrard, A.E., Mohr, J.J., astro-ph/0010584 
\reference{} Blanton, M., Cen, R.,
Ostriker, J.P., Strauss, M.A., \& Tegmark, M. 2000, ApJ, 531, 1
\reference{} Bryan, G.L., \& Norman, M.L. 1998, ApJ, 495, 80
\reference{} Bryan, G.L., astro-ph/0009286 
\reference{} Carlberg, R.G. et al. 1996, ApJ, 462, 32 
\reference{} Cen 1998, ApJ, 509, 494
\reference{} de Bernardis, P. \etal 2000, Nature, 404, 955
\reference{} Eke, V.R., Cole, S., \& Frenk, C.S. 1996, MNRAS, 282, 263 
\reference{} Eke, V.R., Navarro, J.F., Steinmetz, M. 2001, ApJ, 554, 114 
\reference{} Evrard, A.E., Metzler, C.A., \& Navarro, J.F. 1996, ApJ, 469, 494
(EMN) 
\reference{} Finoguenov, A., Reiprich, T. H.,\& Bohringer,
H., astro-ph/0010190 (FRB)
\reference{} Girardi, M.,Giuricin, G.,
Mardirossian, F., Mezzetti, M., Boschin, W. 1998, ApJ, 505, 74
\reference{} Girardi, M., Mezzetti, M. 2000, ApJ, 548, 79
\reference{} Gunn, J., Gott, J. 1972, 176, 1 
\reference{} Henry, J.P. 2000, ApJ, 565, 580 \reference{} Hjorth, J., Oukbir, J. \& van Kampen, E., 1998,MNRAS, 298, L1 
\reference{} Horner, D.J., Mushotzky, R.F., \& Scharf, C.A. 1999, ApJ, 520, 78(HMS)
\reference{} Iliev, I.T., \& Shapiro, P.R. 2001, MNRAS, in press (astro-ph/0101067)
\reference{} Mathiesen, B.F., astro-ph/0012117
\reference{} Mathiesen, B.F., Evrard, A.E. 2001, ApJ 546, 1, 100 
\reference{} Muanwong et al., astro-ph/0102048 \reference{} Navarro, Frenk \& White 1997, ApJ 490, 493(NFW) 
\reference{} Nevalainen, J.,
Markevitch, M., \& Forman, W. 2000, ApJ, 532, 694 
\reference{} Neumann, D.M., \& Arnaud, M. 1999, A\& A, 348, 711 
\reference{} Oukbir, J., Bartlett, J.G., \& Blanchard, A. 1997, A\&A, 320, 365
\reference{} Padmanabhan, T., Structure Formation in the Universe,
1993, Cambridge University Press 
\reference{} Peebles, P.J.E., Daly, R.A., \& Juszkiewicz, R. 1989, \apj, 347, 563 
\reference{} Pen, U. 1998, ApJ, 498, 60
\reference{} Riess, A.G., \etal 1998, AJ, 116, 1009
\reference{} Schindler, S. 1999, A\&A, 349, 435
\reference{} Shapiro, P.R., Iliev, I.T., \& Raga, A.C. 1999, MNRAS, 307, 203
\reference{} Thomas, P.A. et al., astro-ph/0007348
\reference{} Tyson, J.A., Kochanski, G.P., \& dell'Antonio, I.P. 1998, ApJ, 498, L107
\reference{} Voit, M. 2000, ApJ 543, 1, 113
\reference{} Viana, P.T.P, \& Liddle, A.R. 1996, MNRAS, 281, 323
\reference{} White, S.D.M., Efstathiou, G. ,\& Frenk, C.S. 1993, MNRAS, 262, 102
\reference{} Wu, J.-H. P. 2000, astro-ph/0012207
\reference{} Xu, H., Jin, G., Xiang-Ping, W., astro-ph/ 0101564


\end{references}
\end{document}